\documentclass[12pt]{article}
\usepackage{graphics}
\usepackage{amsthm}
\usepackage{amssymb}
\usepackage{amsmath}
\usepackage{amsfonts}
\usepackage{epic}
\usepackage{hyperref}
\usepackage{epsfig}
\usepackage{bm}
\usepackage{amscd}

\theoremstyle{plain}

\theoremstyle{definition}

\def\be{\begin{equation}}
\def\ee{\end{equation}}
\smallskip
\input{tcilatex}

\begin{document}

\headsep=-0.5cm

\begin{titlepage}
\begin{flushright}
\end{flushright}
\begin{center}
\noindent{{\LARGE{A twisted FZZ-like dual for the 2D back hole}}}

\smallskip
\smallskip
\smallskip


\smallskip
\smallskip

\smallskip

\smallskip

\smallskip
\smallskip
\noindent{\large{Gaston Giribet${}^{1,2,3}$ and Mat\'{\i}as Leoni${}^{2}$}}
\end{center}

\smallskip

\smallskip
\smallskip
\centerline{${}^{1}$ Abdus Salam International Centre for Theoretical Pysics, ICTP,}
\centerline{{\it Strada Costiera 11, 34014, Trieste, Italy.}}
\smallskip

\smallskip
\centerline{${}^{2}$ Physics Department, Universidad de Buenos Aires, FCEN - UBA,}
\centerline{{\it Ciudad Universitaria, pabell\'on 1, 1428, Buenos Aires, Argentina.}}
\smallskip
\smallskip
\centerline{${}^{3}$ Consejo Nacional de Investigaciones Cient\'{\i}ficas y T\'ecnicas, CONICET,}
\centerline{{\it Av. Rivadavia, 1917, 1033, Buenos Aires, Argentina.}}
\smallskip
\smallskip

\smallskip
\smallskip

\smallskip

\smallskip

\begin{abstract}
We review and study the duality between string theory formulated on a curved exact background (the two dimensional black
hole) and string theory in flat space with a tachyon-like potential. We generalize previous results in this subject by discussing a
twisted version of the Fateev-Zamolodchikov-Zamolodchikov conjecture. This duality is shown to hold at the level of $N$-point correlation
functions on the sphere topology, and connects tree-level string amplitudes in the euclidean version of the 2D black hole ($\times$ time) to
correlation functions in a non-linear $\sigma $-model in flat space but in presence of a tachyon wall potential and a linear dilaton.
The dual CFT corresponds to the perturbed 2D quantum gravity coupled to $c<1$ matter ($\times$ time), where the operator that describes the tachyon-like potential
can be seen as a $n=2$ momentum mode perturbation, while the usual sine-Liouville
operator would correspond to the vortex sector $n =1$. We show how the sine-Liouville interaction term arises through a twisting of the marginal deformation introduced here, and discuss such 'twisting' as a non-trivial realization
of the symmetries of the theory. After briefly reviewing the computation of correlation functions in sine-Liouville
CFT, we give a precise prescription for computing correlation functions in the twisted model. To show the new version of the correspondence we make use of a formula recently proven by S. Ribault and J. Teschner, which connects the correlation functions in the Wess-Zumino-Witten
theory to correlation functions in the Liouville theory. Conversely, the duality discussed here can be thought of as a free field realization of such remarkable formula.
\[
\]
This paper is an extended version of the authors' contribution to the XVIth International Colloquium on Integrable 
Systems and Quantum Symmetries, held in Prague, Czech Republic, in June 2007. A brief version was published
in Rep. Math. Phys. 61 2 (2008) 151-162. Part of the material presented here is based on the results that one of the 
authors has reported in Refs. \cite{0511252,gaston}, and it is in some way related to the recent works \cite{RT,R,MMP,Nuevo,Fateev}.

\end{abstract}

\end{titlepage}



\newpage


\section{Introduction}

One of the most profound concepts in string theory is the suggestive idea
that spacetime itself could be a mere emergent notion, a sort of effective
description of a more fundamental entity \cite{Seiberg}. This conception
relies on the existence of the duality symmetries of string theory, which
suggest that concepts such as the curvature and topology of the spacetime
might be only auxiliary notions. This idea is particularly realized by
examples that manifestly show the duality between string theory formulated
on curved backgrounds (e.g. black holes) and the theory in flat space but in
presence of tachyon-like potentials. This is the subject we will explore
here; and we will do this by studying the worldsheet description of the 2D
string theory in the black hole background (i.e. the gauged $SL(2,\mathbb{R}%
)_{k}/U(1)$ Wess-Zumino-Witten (WZW) model).

\subsubsection*{1.1 \ The subject}

The relation between string theory in the 2D black hole background and
Liouville-like conformal field theories representing \textquotedblleft
tachyon wall" potentials was extensively explored in the past. One of the
celebrated examples is the Mukhi-Vafa duality \cite{MV}, relating a twisted
version of the euclidean black hole to the $c=1$ matter coupled to 2D
gravity. The literature on the connection between the $c=1$ CFT and the
black hole CFT is actually quite rich; we should refer to the list of papers 
\cite{22}-\cite{Sameer} and the references therein. Recently, a new relation
between the 2D string theory in the euclidean black hole background and a
deformation of the $c=1$ matter CFT has received remarkable attention: This
is the so-called Fateev-Zamolodchikov-Zamolodchikov conjecture (FZZ), which
states the equivalence between the black hole and the often called
sine-Liouville field theory \cite{FZZ,KKK}. In the last six years this FZZ
duality has been applied to study the spectrum and interactions of strings
in both the black hole geometry and the Anti-de Sitter space \cite%
{Notes,YoYu,Takayanagi2}; and the most important application of it was so
far the formulation of the matrix model for the two-dimensional black hole 
\cite{KKK}. In fact, when one talks about the \textquotedblleft black hole
matrix model" one is actually referring to the matrix model for the
sine-Liouville deformation of the $c=1$ matter CFT, and thus the black hole
description in such a framework emerges through the FZZ correspondence. This
manifestly shows how useful the FZZ duality is in the context of string
theory.

Although at the beginning it appeared as a conjecture, a proof of the FZZ
duality was eventually given some years ago\footnote{%
More recently, after this paper was published, Y. Hikida and V. Schomerus
presented a proof of the FZZ conjecture \cite{HikidaSchomerus2}.}. This was
done in two steps: first, by proving the equivalence of the corresponding
N=2 supersymmetric extensions of both the 2D black hole $\sigma $-model and
the sine-Liouville theory \cite{HK}; and, secondly, by showing that the
fermionic parts of the N=2 theories eventually decouple, yielding the
bosonic duality as an hereditary property \cite{Maldacena}, see also \cite%
{IPT,IPT2}. This could be done because both sine-Liouville and the black
hole theory admit a natural\footnote{%
The 2D black hole can be realized by means of the Kazama-Suzuki construction 
\cite{KS,KS2}, while the sine-Liouville theory can be seen as a sector of
the N=2 Liouville theory. The bosonic version of the FZZ duality can be seen
to arise by GKO quotienting the $U(1)$ R-symmetry of the N=2 version.}
embedding in N=2 theories, where the duality can be seen as a manifestation
of the mirror symmetry. However, one could be also interested in seeing
whether a proof of such a duality exists at the level of the bosonic theory
itself. In this paper we will show how such a duality can be actually proven
(at the level of the sphere topology) without resorting to arguments based
on supersymmetry but just making use of the conformal structure of the
theory.

\subsubsection*{1.2 \ The result}

We will show that any $N$-point correlation functions in the $SL(2,\mathbb{R}%
)_{k}/U(1)$ WZW ($\times $ $time$) on the sphere topology is equivalent to a 
$N$-point correlation functions in a two-dimensional conformal field theory
that describes a linear dilaton $\sigma $-model perturbed by a tachyon-like
potential. This actually resembles the FZZ correspondence; however, instead
of considering a vortex perturbation with winding $|n|=1$ here we will
consider momentum modes of the sector $n=2$. To be precise, the theory we
will consider is defined by turning on the modes $\lambda _{n=2}\neq 0$ and $%
\lambda _{n=1}\neq 0$ in the following action 
\begin{equation}
S=\frac{1}{4\pi }\int d^{2}z\left( \partial X\overline{\partial }X+\partial
\varphi \overline{\partial }\varphi -\frac{1}{2\sqrt{2}}\widehat{Q}R\varphi
+\sum_{n}\lambda _{n}e^{-\frac{\alpha _{n}}{\sqrt{2}}\varphi +in\sqrt{\frac{k%
}{2}}X}\right)   \label{S}
\end{equation}%
where $\widehat{Q}=(k-2)^{-1/2}$ and $\alpha _{n}=\widehat{Q}(1+\sqrt{%
1+(kn^{2}-4)(k-2)})$. Namely, the perturbation we will consider is given by
the operator%
\begin{equation}
\mathcal{O}=\lambda _{1}e^{-\sqrt{\frac{k-2}{2}}\varphi +i\sqrt{\frac{k}{2}}%
X}+\lambda _{2}e^{-\sqrt{\frac{2}{k-2}}(k-1)\varphi +i\sqrt{2k}X},
\label{Pupapupapupa56}
\end{equation}%
where we denoted $X=X_{L}(z)+X_{R}(\overline{z})$, which has to be
distinguished from the T-dual direction $\widetilde{X}=X_{L}(z)-X_{R}(%
\overline{z})$. Operators $e^{-\frac{\alpha _{n}}{\sqrt{2}}\varphi +in\sqrt{%
\frac{k}{2}}X}$ are ($1,1$)-operators with respect to the stress-tensor of
the free theory%
\begin{equation}
T(z)=-\frac{1}{2}\left( \partial X\right) ^{2}-\frac{1}{2}\left( \partial
\varphi \right) ^{2}-\frac{\widehat{Q}}{\sqrt{2}}\partial ^{2}\varphi ,
\label{ThisCondition}
\end{equation}%
so that they represent marginal deformations of the linear dilaton theory.
However, it is worth pointing out that condition (\ref{ThisCondition}) is
not sufficient to affirm that the theory defined by action (\ref{S}) is 
\textit{exactly} marginal. In general, proving a theory is an exact
conformal field theory is highly non-trivial. Nevertheless, there is strong
evidence that particular perturbations belonging to those in (\ref{S}) do
represent\footnote{%
One example of such a perturbation is sine-Liouville potential, which we
will discuss in section 3. Notice also that, at the critical value $k=9/4$,
the perturbations in (\ref{S}) are precisely those discussed in \cite{KKK}
in the context of matrix model.} CFTs.

Coefficients $\lambda _{n}$ in (\ref{S}) must satisfy the condition $\lambda
_{n}=\lambda _{-n}$ for the Lagrangian to be real, and thus the theory
results invariant under $X\rightarrow -X$. The scaling relations between
different couplings $\lambda _{n}$ are given by standard KPZ arguments \cite%
{KPZ,David,KPZ2}, being the scale of the theory governed by one of these
constants, analogously as to how the Liouville cosmological constant
introduces the scale in the $c=1$ matter CFT. The central charge of the
theory is then obtained from the operator product expansion of the
stress-tensor, yielding $c=2+6\widehat{Q}^{2}=2+\frac{6}{k-2}.$ Eventually,
we will be interested in adding a time-like free boson to the theory in
order to define a Lorentzian target space of the form $SL(2,\mathbb{R}%
)_{k}/U(1)\times time$, so the central charge will receive an additional
contribution $+1$ coming from the time $\mathbb{R}$ direction, yielding 
\begin{equation}
c=3+6\widehat{Q}^{2}=3+\frac{6}{k-2},  \label{c}
\end{equation}%
while the stress-tensor will result supplemented by a term $+\frac{1}{2}%
\left( \partial T\right) ^{2}$. For practical purposes, this time-like
direction can be thought of as an auxiliary degree of freedom, and it does
not enter in the non-trivial part of the duality we want to discuss, being
coupled to the other directions just by the value of the central charge%
\footnote{%
In the case the theory corresponds to the product $SL(2,\mathbb{R}%
)/U(1)\times time$ the condition $c=26$ demands $k=52/23$. On the other
hand, if the space is just the coset $SL(2,\mathbb{R})/U(1)$ the
corresponding condition reads $k=9/4$.} $c$.

\subsubsection*{1.3 \ Outline}

The particular correspondence between the model (\ref{S}) and the 2D black
hole we will discuss turns out to be realized at the level of $N$-point
functions on the sphere topology, and corresponds to a twisted version of
the FZZ correspondence\footnote{%
In the sense that it involves a deformation of the sine-Liouville
interaction term in the action.}. Consequently, we will discuss the latter
first. While being similar, the duality we will discuss herein presents two
important differences with respects to the FZZ: The first difference is that
the new duality admits to be proven\footnote{%
cf. Ref. \cite{HikidaSchomerus2}.} in a relatively simple way without
resorting to arguments based on mirror symmetry of its supersymmetric
extension; secondly, it involves higher momentum modes ($n=2$) instead of
winding modes of the sector $n=1$. We will make the precise statement of the
new correspondence in section 4, where we also address its proof. The paper
is organized as follows: In section 2 we review some features of the
conformal field theories that play an important role in our work. First, we
review the computation of correlation functions in Liouville field theory
with the purpose of emphasizing some features and refer to the analogy with
the Liouville case whenever an illustrative example is needed. Secondly, we
discuss some general aspects of the 2D black hole $\sigma $-model. Once
these two CFTs are introduced, we discuss how correlation functions in both
theories are related through a formula recently proven by S. Ribault and J.
Teschner \cite{RT,R}. Their formula connects correlation functions in both
WZW and Liouville theory in a remarkably direct way \cite{S}, and it turns
out to be important for proving our result. In section 3 we briefly review
the FZZ dual for the 2D black hole; namely the sine-Liouville field theory.
In section 4 we introduce a \textquotedblleft twisted" version of the
sine-Liouville theory, and we show that such \textquotedblleft deformed"
sine-Liouville turns out to be a dual for the 2D black hole as well. A
crucial piece to show this new version of the duality is the
Ribault-Teschner formula mentioned above, for which we present a free field
realization that is eventually identified as being precisely the deformed
sine-Liouville model we want to study. Section 5 contains the conclusions.

\section{Conformal field theory}

To begin with, let us discuss some aspects of correlation functions in
Liouville field theory. The reason for doing this is that Liouville theory
is the prototypical example of non-compact conformal field theory \cite%
{Schomerus} and thus the techniques for computing correlation functions in
this model are analogous to those we will employ in the rest of the paper.
Moreover, the models we will consider here are actually deformations of the
Liouville theory coupled to a $c=1+1$ matter field, so that it is clearly
convenient to consider this model first.

\subsection{Liouville theory}

\subsubsection{Liouville field theory coupled to $c=1(+1)$ matter}

Liouville theory naturally arises in the formulation of the two-dimensional
quantum gravity and in the path integral quantization of string theory \cite%
{Polyakov}. This is a non-trivial conformal field theory \cite{Yu,Teschner}\
whose action reads%
\begin{equation}
S_{L}[\mu ]=\frac{1}{4\pi }\int d^{2}z\left( \partial \varphi \overline{%
\partial }\varphi +\frac{1}{2\sqrt{2}}QR\varphi +4\pi \mu e^{\sqrt{2}%
b\varphi }\right)  \label{mancha}
\end{equation}%
where $\mu $ is a real positive parameter called \textquotedblleft the
Liouville cosmological constant". The background charge parameter takes the
value $Q=b+b^{-1}$ in order to make the Liouville barrier potential $\mu e^{%
\sqrt{2}b\varphi }$ to be a marginal operator. In the conformal gauge, the
linear dilaton term $QR\varphi $, which involves the two-dimensional Ricci
scalar $R,$ has to be understood as keeping track of the coupling with the
worldsheet curvature that receives a contribution coming from the point at
infinity. The theory is globally defined once one specifies the boundary
conditions, and this can be done by imposing the behavior $\varphi \sim -2%
\sqrt{2}Q\log |z|$ for large $|z|$, that is compatible with the spherical
topology. Under holomorphic transformations $z\rightarrow w$ Liouville field
transforms in a way that depends on $Q$, namely $\varphi \rightarrow \varphi
-\sqrt{2}Q\log |\frac{dw}{dz}|$. In this paper we will be interested in the
coupling of Liouville theory to a $U(1)$ boson field represented by an
additional $-\frac{1}{4\pi }\int d^{2}z\partial X\overline{\partial }X$
piece in the action (\ref{mancha}) above. Moreover, we can also include the
\textquotedblleft time" direction $\frac{1}{4\pi }\int d^{2}z\partial T%
\overline{\partial }T.$ Then, the central charge of the whole theory is
given by%
\begin{equation*}
c=2+c_{L}=3+6Q^{2},
\end{equation*}%
where $c_{L}$ refers to the Liouville central charge. Important objects of
the theory are the exponential vertex operators \cite{Teschnervertex} 
\begin{equation*}
V_{\alpha }(z)\times e^{i\sqrt{2}p_{1}X(z)+i\sqrt{2}p_{0}T(z)}=e^{\sqrt{2}%
\alpha \varphi (z)+i\sqrt{2}p_{1}X(z)+i\sqrt{2}p_{0}T(z)},
\end{equation*}%
which turn out to be local operators of conformal dimension $h=\alpha
(Q-\alpha )+p_{1}^{2}-p_{0}^{2}$ with respect to the stress-tensor $T(z)$ of
the free theory,%
\begin{equation}
T(z)=\frac{1}{2}(\partial T)^{2}-\frac{1}{2}(\partial X)^{2}-\frac{1}{2}%
(\partial \varphi )^{2}+\frac{Q}{\sqrt{2}}\partial ^{2}\varphi .  \label{T}
\end{equation}

Now, let us move on and discuss correlation functions.

\subsubsection{Liouville correlation functions}

The non-trivial part of correlation functions in the theory (\ref{T}) is
given by the Liouville correlation functions \cite{Teschner,ZZ,Zreloaded,DO}%
, and these are formally defined as follows%
\begin{equation*}
A_{(\alpha _{1},...\alpha _{N}|z_{1},...z_{N})}^{L}=\left\langle V_{\alpha
_{1}}(z_{1})...V_{\alpha _{N}}(z_{N})\right\rangle _{S_{L}[\mu ]}=\int
D\varphi e^{-S_{L}[\mu ]}\prod_{i=1}^{N}e^{\sqrt{2}\alpha _{i}\varphi
(z_{i})}
\end{equation*}%
and, on the spherical topology, these can be written by using that 
\begin{align}
\left\langle \prod_{i=1}^{N}V_{\alpha _{i}}(z_{i})\right\rangle _{S_{L}[\mu
]}& =b^{-1}\mu ^{s}\Gamma (-s)\delta \left( s+b^{-1}(\alpha _{1}+\alpha
_{2}+...\alpha _{N})-1-b^{-2}\right) \times  \notag \\
& \times \prod_{r=1}^{s}\int d^{2}w_{r}\left\langle \prod_{i=1}^{N}V_{\alpha
_{i}}(z_{i})\prod_{r=1}^{s}V_{b}(w_{r})\right\rangle _{S_{L}[\mu =0]},
\label{despues}
\end{align}%
namely,%
\begin{equation}
A_{(\alpha _{1},...\alpha _{N}|z_{1},...z_{N})}^{L}=b^{-1}\mu ^{s}\Gamma
(-s)\delta \left( s+b^{-1}(\alpha _{1}+\alpha _{2}+...\alpha
_{N})-1-b^{-2}\right) \times  \notag
\end{equation}%
\begin{equation}
\times \prod_{r=1}^{s}\int d^{2}w_{r}\int D\varphi e^{-S_{L}[\mu
=0]}\prod_{i=1}^{N}e^{\sqrt{2}\alpha _{i}\varphi (z_{i})}\prod_{r=1}^{s}e^{%
\sqrt{2}b\varphi (w_{r})}.  \label{da}
\end{equation}%
This permits to compute correlation functions by employing the standard
Gaussian measure and free field techniques. The overall factor $\Gamma (-s)$
and the $\delta $-function come from the integration over the zero-mode $%
\varphi _{0}$ of the Liouville field $\varphi $, and it also yields the
insertion of an specific amount, $s,$ of screening operators $V_{b}(w)$ in
the correlator. In deriving (\ref{da}), the identity $\mu ^{s}\Gamma
(-s)=\int dxx^{-1-s}e^{-\mu x}$ and the Gauss-Bonnet theorem were used to
find out the relation between $s$, $b$, and the momenta $\alpha _{i}$, which
for a manifold of generic genus $g$ and $N$ punctures would yield%
\begin{equation}
bs+\sum_{i=1}^{N}\alpha _{i}=Q(1-g).  \label{pupo2}
\end{equation}%
So, the correlators can be computed through the Wick contraction of the $N+s$
operators by using the propagator $\left\langle \varphi (z_{1})\varphi
(z_{2})\right\rangle =-2\log |z_{1}-z_{2}|,$ which corresponds to the free
theory (\ref{T}) and yields the operator product expansion $e^{\alpha
_{1}\varphi (z_{1})}e^{\alpha _{2}\varphi (z_{2})}\sim
|z_{1}-z_{2}|^{-2\alpha _{1}\alpha _{2}}e^{(\alpha _{1}+\alpha _{2})\varphi
(z_{2})}+...$. In principle, this could be used to integrate the expression
for $A_{(\alpha _{1},...\alpha _{N}|z_{1},...z_{N})}^{L}$ explicitly.
Nevertheless, it is worth noticing that the expression (\ref{da}) can be
considered just formally since, in general, $s$ is not an integer number.
Hence, in order to compute generic correlation functions one has to deal
with the problem of making sense of such integral representation. With the
purpose of giving an example, let us describe below the computation of the
partition function on the sphere in detail. Such case corresponds to $g=0$
and $N=0$, and the number of screening operators to be integrated out turns
out to be $m=s-3=-2+b^{2}$. That is, in order to compute the genus zero
partition function we have to consider the correlation function of three
local operators $e^{\sqrt{2}b\varphi (z)}$ inserted at the points $%
z_{1}=0,z_{2}=1$ and $z_{3}=\infty $ to compensate the volume of the
conformal Killing group, $SL(2,\mathbb{C})$. This has to be distinguished
from the direct computation of the three-point function \cite{ZZ} of three
\textquotedblleft light" states $\alpha _{1}=\alpha _{2}=\alpha _{3}=b,$ as
we will discuss below.

\subsubsection{A working example: the spherical partition function}

Although it is usually said that string partition function on the spherical
topology vanishes, we know that this is not necessary the case when the
theory is formulated on non-trivial backgrounds. A classical example of this
is the two-dimensional string theory formulated in both tachyonic and
gravitational non-trivial backgrounds we will be discussing along this
paper. Such models admit a description in terms of the Liouville-type sigma
model actions, so that the computation of the corresponding genus zero
partition functions involves the computation of spherical partition function
of Liouville theory or some deformation of it. Here, we will describe a
remarkably simple calculation of the Liouville partition function on the
spherical topology by using the free field techniques. The free field
techniques to be employed here were developed so far by Dotsenko and Fateev 
\cite{Fateev2,Dotsenko}, and by Goulian and Li \cite{GLi} (see also \cite%
{dot1,dot2,D}). The partition function $Z_{g=0}$ is then given by%
\begin{equation}
Z_{g=0}=\frac{\mu ^{m+3}}{b}\Gamma (-m-3)\lim_{z_{3}\rightarrow \infty
}|z_{3}|^{-4}\prod_{r=1}^{m}\int d^{2}w_{r}\int D\varphi e^{-S_{L}[\mu
=0]}e^{\sqrt{2}b\varphi (0)}e^{\sqrt{2}b\varphi (1)}e^{\sqrt{2}b\varphi
(z_{3})}\prod_{r=1}^{m}e^{\sqrt{2}b\varphi (w_{r})}.  \label{empieza}
\end{equation}%
with $m=-2+b^{-2}$. According to the standard Wick rules, we can write%
\begin{equation*}
Z_{g=0}=b^{-1}\mu ^{3+m}\Gamma (-m-3)\prod_{r=1}^{m}\int d^{2}w_{r}\left(
\prod_{r=1}^{m}|w_{r}|^{4\rho }|1-w_{r}|^{4\rho
}\prod_{r<t}^{m-1,m}|w_{t}-w_{r}|^{4\rho }\right) .
\end{equation*}%
This can be explicitly solved for integer $m$ by using the Dotsenko-Fateev
integral formula worked out in reference \cite{Dotsenko}. Even though we are
interested in the case where $m$ is generic enough, and this can mean a
negative real number, we can assume that this is an integer positive number
through the integration and then try to analytically extend the final
expression accordingly. In this way, we get%
\begin{equation*}
Z_{g=0}=\frac{\mu ^{3+m}}{b}\Gamma (-m-3)\Gamma (m+1)\pi ^{m}\gamma
^{m}(1-\rho )\prod_{r=1}^{m}\gamma (r\rho )\prod_{r=0}^{m-1}\gamma
^{2}(1+(2+r)\rho )\gamma (-1-(3+r+m)\rho ).
\end{equation*}%
where, as usual, we denoted $\gamma (x)=\Gamma (x)/\Gamma (1-x)$; and we
also denoted $\rho =-b^{2}$ for notational convenience. Once again, this
expression only makes sense for $m$ being a positive integer number, so that
the non-trivial point here is that of performing analytic continuation. In
order to do this, we can rewrite the expression above by taking into account
that $\gamma (-1-(3+r+m)\rho )=\gamma (-(r+1)\rho ).$ So we can expand it as%
\begin{equation}
Z_{g=0}=\frac{\mu ^{3+m}}{b}\Gamma (-m-3)\Gamma (m+1)\pi ^{m}\gamma
^{m}(1-\rho )\prod_{r=1}^{m}\gamma (r\rho )\gamma (-r\rho
)\prod_{r=2}^{m+1}\gamma ^{2}(1+r\rho ).  \label{introducidas}
\end{equation}%
Now, some simplifications are required. First, we can use that $%
m=-2+b^{-2}=-2-\rho ^{-1}$ and $1+r\rho =-(m+2-r)\rho $ to arrange the last
product. Then, we can rewrite the product as%
\begin{equation*}
\gamma (1+2\rho )\gamma (1+3\rho )...\gamma (1+m\rho )\gamma (1+(m+1)\rho
)=\gamma (-\rho )\gamma (-2\rho )...\gamma (-(m-1)\rho )\gamma (-m\rho ),
\end{equation*}%
that is%
\begin{equation*}
\prod_{r=2}^{m+1}\gamma (1+r\rho )=\prod_{r=1}^{m}\gamma (-r\rho ),
\end{equation*}%
and then use $\gamma (r\rho )\gamma (1-r\rho )=1$ to write%
\begin{equation*}
Z_{g=0}=b^{-1}\mu ^{Q/b}\Gamma (-m-3)\Gamma (m+1)\pi ^{m}\gamma ^{m}(1-\rho
)\gamma ^{2}(-\rho )(-1)^{m}\rho ^{-2m}\Gamma ^{-2}(m+1),
\end{equation*}%
where the identities $\gamma (x)\gamma (-x)=\gamma (x)/\gamma (1+x)=-x^{-2}$
were also used. Again, the properties of the $\gamma $-function can be used
to write $\gamma (2+\rho ^{-1})=-(1+\rho ^{-1})^{2}\gamma (1+\rho ^{-1}),$ $%
\gamma (1-\rho )=-\rho ^{2}\gamma (-\rho )$ and $\gamma (-1-\rho )=-(1+\rho
)^{-2}\gamma (-\rho )$. Then, once all is written in terms of $b,$ the
partition function reads\footnote{%
Notice that we have absorbed a factor $\sqrt{2}$ in the definition of the
measure of the path integral.}%
\begin{equation}
Z_{g=0}=\frac{(1-b^{2})\left( \pi \mu \gamma (b^{2})\right) ^{Q/b}}{\pi
^{3}Q\gamma (b^{2})\gamma (b^{-2})}.  \label{ZZZ}
\end{equation}%
This is the exact result for the Liouville partition function on the
spherical topology, which turns out to be a non trivial function of $b$. It
oscillates with growing frequency and decreasing amplitude according $b^{2}$
approaches the values $b^{2}=0$ and $b^{2}=1$. One of the puzzling features
of the expression (\ref{ZZZ}) is the fact that it does not manifest the
self-duality that the Liouville theory seems to present under the
transformation $b\rightarrow 1/b$. In order to understand this point, it is
convenient to compare the direct computation of $Z_{g=0}$ we gave above with
the analogous computation of the Liouville structure constant (three-point
functions) $C(\alpha _{1},\alpha _{2},\alpha _{3})$ for the particular
configuration $\alpha _{1}=\alpha _{2}=\alpha _{3}=b$. The difference
between both calculations is given by the overall factor $\Gamma (-s)=\Gamma
(-m-3)$ in (\ref{empieza}). As mentioned, this factor comes from the
integration over the zero-mode of the field $\varphi $, but it can be also
thought of as coming from the combinatorial problem of permuting all the
screening operators. Actually, for integer $s$ this factor can be written as 
$\Gamma (-s)=(-1)^{s}\Gamma (0)/s!$, where the divergent factor $\Gamma (0)$
keeps track of a divergence due to the non-compactness of the Liouville
direction. In fact, this yields the factorial $1/s!$ arising in the residue
corresponding to the poles of resonant correlators. On the other hand, in
the case of being computing the structure constant $C(b,b,b)$, unlike the
computation of $Z_{g=0}$, such overall factor should be $\Gamma (3-s)$
instead of $\Gamma (-s)$ since one has to divide by the permutation of $s-3$
screening charges. Hence, we have $C(b,b,b)/Z_{g=0}=\Gamma (3-s)/\Gamma
(-s)=-s!/(s-3)!=-(b^{-2}+1)b^{-2}(b^{-2}-1)$. This is precisely consistent
with the fact that $\frac{d^{3}Z}{d\mu ^{3}}=-C(b,b,b)\sim \mu ^{Q/b-3}$,
see Ref. \cite{Z}. Thus, this combinatorial problem appears as being the
origin of the breakdown of the Liouville self-duality at the level of the
partition function.

Now, let us move to study another CFT that is also a crucial piece in our
discussion: the CFT that describes the 2D black hole $\sigma $-model.

\subsection{String theory in the 2D black hole}

\subsubsection{The action and the semiclassical picture}

String theory in two dimensions presents very interesting properties that
make of it a fruitful ground to study features of its higher dimensional
analogues. One example is given by the 2D black hole solution discovered in
Refs. \cite{Witten,Maldal,Pakmandice}. This black hole solution is supported
by a dilaton configuration, and it turns out to be an exact conformal
background on which formulate string theory. In fact, the 2D black hole $%
\sigma $-model action corresponds to the gauged level-$k$ $SL(2,\mathbb{R}%
)_{k}/U(1)$ WZW theory \cite{Witten}. An excellent comprehensive review on
this model can be found in Ref. \cite{Persson}.

The worldsheet action for string theory in a two-dimensional metric-dilaton
background, once setting $\alpha ^{\prime }=2$, reads%
\begin{equation}
S_{P}=\frac{1}{4\pi }\int d^{2}z\left( G_{\mu \nu }(X)\partial X^{\mu }%
\overline{\partial }X^{\nu }+R\Phi (X)\right) ,  \label{Polyakov2}
\end{equation}%
where the indices $\mu ,\nu =\{1,D=2\}$ run over the two coordinates of the
target space, whose metric is $G_{\mu \nu }(X)$. This action is written in
the conformal gauge, so, as we discussed before, the dilaton term $R$%
\thinspace $\Phi (X)$ has to be understood as keeping track of the coupling
with the worldsheet curvature that receives a contribution coming from the
point at infinity. The vanishing of the one-loop $\beta $-functions demands $%
R_{\mu \nu }=\nabla _{\mu }\nabla _{\nu }\Phi $, with $R_{\mu \nu }$ being
now the Ricci tensor associated to the target space metric $G_{\mu \nu }$.
Since the 2D black hole string theory corresponds to the $SL(2,\mathbb{R}%
)_{k}/U(1)$ WZW model, it admits an exact algebraic description in terms of
the current conformal algebra of the WZW theory; and we will comment on this
in the following subsection. In the semiclassical limit, governed by the
large $k$ regime, the euclidean version of the background is described by
the following configurations for the metric $G_{\mu \nu }$ and the dilaton $%
\Phi $,%
\begin{equation*}
ds^{2}=k\left( dr^{2}+\tanh ^{2}r\ dX^{2}\right) ,\quad \quad \Phi (r)=\Phi
_{0}-2\log \left( \cosh r\right) .
\end{equation*}%
It is well known that the geometry of the euclidean black hole is that of a
semi-infinite cigar that asymptotically looks like a cylinder. The angular
coordinate of such cylinder is $X$, while the coordinate $r$ is the one that
goes along the cigar, running from $r=0$ (the tip of the cigar, where the
string theory is strongly coupled) to $r=\infty $ (where the string coupling 
$e^{\Phi (r)}$ tends to zero). To get a semiclassical picture of this
geometry, let us consider the large $k$ regime and redefine the radial
coordinate as $\cosh ^{2}r=M^{-1}e^{\sqrt{2/k}\varphi }$. Then, in the large 
$\varphi $ approximation, and by also rescaling the angular coordinate $X$
by a factor $\sqrt{2/k},$ the metric reads%
\begin{equation}
ds^{2}=2\left( 1+Me^{-\sqrt{2/k}\varphi }\right) d\varphi ^{2}+2\left(
1-Me^{-\sqrt{2/k}\varphi }\right) dX^{2},  \label{H}
\end{equation}%
that asymptotically looks like the cylinder of radius $R=\sqrt{k/2}$. The
parameter $M$ is related to the mass of the black hole, and it can be fixed
to any positive value by shifting $\varphi $. Considering finite-$k$
corrections leads to a shifting in $k$ and then the metric and the dilaton
result corrected. In such case, the dilaton reads%
\begin{equation*}
\Phi (\varphi )=\Phi _{0}-\log M+\sqrt{2}\widehat{Q}\varphi ,\qquad \widehat{%
Q}=(k-2)^{-1/2}.
\end{equation*}%
Thus, the 2D string theory in the euclidean black hole background can be
semiclassically described by a deformation of the linear dilaton theory 
\begin{equation}
S_{0}=\frac{1}{4\pi }\int d^{2}z\left( \partial X\overline{\partial }%
X+\partial \varphi \overline{\partial }\varphi -\frac{1}{2\sqrt{2}}\widehat{Q%
}R\varphi \right) ;  \label{S0}
\end{equation}%
and, according to (\ref{H}) and taking into account the finite-$k$
corrections, such \textquotedblleft deformation" corresponds to perturbing
the action (\ref{S0}) with the graviton-like operator \cite{24}%
\begin{equation}
\mathcal{O}=M\ \partial X\overline{\partial }X\ e^{-\sqrt{\frac{2}{k-2}}%
\varphi };  \label{graviton}
\end{equation}%
this is true up to a BRST-trivial\footnote{%
That means that it is pure gauge in the BRST cohomology.} operator of the
form $\delta \mathcal{O}\sim \partial \varphi \overline{\partial }\varphi \
e^{-\sqrt{\frac{2}{k-2}}\varphi }.$ In these terms, the theory can be in
principle solved (e.g. its correlation functions can be computed) by using
the free field approach and the Coulomb-like correlators $\left\langle
\varphi (z_{1})\varphi (z_{2})\right\rangle =\left\langle
X(z_{1})X(z_{2})\right\rangle =-2\log |z_{1}-z_{2}|$. Operator (\ref%
{graviton}) is usually called the \textquotedblleft black hole mass
operator". The inclusion of this operator in the action has to be thought of
as being valid in a semiclassical picture and can be shown to be equivalent
to the free field representation of the WZW model.

In the large $\varphi $ region of the space (where the theory turns out to
be weakly coupled) we have that the non-linear $\sigma $-model of strings in
the black hole seems to coincide with the action $S_{0}+\frac{1}{4\pi }\int
d^{2}z\ \mathcal{O}$. Furthermore, there is a way of seeing that operator (%
\ref{graviton}) actually describes the dilatonic black hole $\sigma $-model
beyond the semiclassical picture. To do so, it is necessary to argue that
such an action unambiguously describes the full theory beyond the weak limit
region \cite{Mukhi,MMP} and, for instance, reproduces the exact correlation
functions. This seems to be hard to be proven in general; nevertheless,
there is a nice way of showing that the perturbation (\ref{graviton})
corresponds to the theory on the black hole background. This relies on the
algebraic description of the $SL(2,\mathbb{R})_{k}/U(1)\times \mathbb{R}$
WZW theory and is quite direct: The point is that the action $S_{0}+\frac{1}{%
4\pi }\int d^{2}z\ \mathcal{O}$ , once supplemented with the BRST-trivial
operator $\delta \mathcal{O}$ and a free time-like boson $-\frac{1}{4\pi }%
\int d^{2}z\ \partial T\overline{\partial }T$, can be shown to be related to
the well known free field realization of the $SL(2,\mathbb{R})_{k}$ WZW
action through a $SO(2,1)$-boost given by\footnote{%
Please, do not mistake the time-like coordinate $T$ for the notation used
for the stress-tensor. Excuse us for this overlap in the notation.} 
\begin{equation*}
T=i\sqrt{\frac{2}{k}}u-i\sqrt{\frac{k-2}{k}}\phi ,\ \ \ \ X=-\sqrt{\frac{k}{2%
}}v+i\frac{k-2}{\sqrt{2k}}u+i\sqrt{\frac{k-2}{k}}\phi ,\ \ \ \ \varphi =%
\sqrt{\frac{k-2}{2}}(u+iv)+\phi ,
\end{equation*}%
and the standard bosonization\footnote{%
It is usually convenient to use a different bosonization, expressing the
field $\beta $ as an exponential function. This would lead to a
Liouville-like interaction in the action.} $\gamma =e^{u+iv}$, $\beta
=i\partial ve^{-u-iv}$, with $\left\langle \beta (z_{1})\gamma
(z_{2})\right\rangle \sim (z_{1}-z_{2})^{-1},$ and with $\left\langle \phi
(z_{1})\phi (z_{2})\right\rangle =-2\log |z_{1}-z_{2}|,$ \cite{HHS}. In
fact, this leads to the Wakimoto free field description of the $SL(2,\mathbb{%
R})_{k}$ current algebra in terms of the linear dilaton field $\phi $ and
the $\beta ,\gamma $ ghost system \cite{Wakimoto}. In Wakimoto variables one
identifies the theory as being the WZW model formulated on $SL(2,\mathbb{R})$
with the elements of the group written in the Gauss parameterization. Then,
the coset theory $SL(2,\mathbb{R})_{k}/U(1)$ is obtained by simply taking
out the time-like direction $T$ which realizes the $U(1)$ current\footnote{%
Alternatively, an additional free boson, analogous to $X,$ can be added in
order to relize the gauging, see \cite{DVV,BB2} and referenctes therein.} 
\begin{equation*}
J^{3}=\beta \gamma +\sqrt{\frac{k-2}{2}}\partial \phi =i\sqrt{\frac{k}{2}}%
\partial T;
\end{equation*}%
recall that this is a time-like direction so that the corresponding
correlator flips its sign and thus turns out to be $\left\langle
T(z_{1})T(z_{2})\right\rangle =+2\log |z_{1}-z_{2}|$.

On the other hand, let us mention that the dual theory (i.e. the
sine-Liouville theory) is also defined as a perturbation of (\ref{S0}); see (%
\ref{cos}) below. According to this picture, it is possible to consider FZZ
duality as a relation between different marginal deformations of the same
free linear dilaton background. This was the philosophy in Ref. \cite{MMP},
where the FZZ correspondence was seen from a generalized perspective,
considering it as an example of a set of connections existing between
different marginal deformations of (\ref{S0}). Here, we will be discussing a
similar correspondence; we will consider perturbations carrying momentum
modes $n=2$ of the tachyon potential and discuss how it describes $SL(2,%
\mathbb{R})_{k}/U(1)\times \mathbb{R}$ WZW correlation functions. We will
dedicate some effort to understand the relation between such $n=2$
perturbation and the standard FZZ duality (that involves $n=1$ modes). But,
first, let us continue our description of the theory in the black hole
background with appropriate detail.

\subsubsection{String spectrum in the 2D black hole and its relation to $%
AdS_{3}$ strings}

The spectrum of the 2D sting theory in the black hole background corresponds
to certain sector of the Hilbert space of the gauged $SL(2,\mathbb{R}%
)_{k}/U(1)$ WZW model, and is thus given in terms of certain representations
of $SL(2,\mathbb{R})_{k}\times \overline{SL}(2,\mathbb{R})_{k}$. The string
states are thus described by vectors $\left| \Phi _{j,m,\bar{m}}^{\omega
}\right\rangle $ which are associated to vertex operators $\Phi _{j,m,\bar{m}%
}^{\omega },$ where $j$, $m,$ and $\bar{m}$ are indices that label the
states of the representations of the group. In order to define the string
theory, it is necessary to identify which is the subset of representations
that have to be taken into account. Such a subset has to satisfy several
requirements\footnote{%
For an interesting discussion on non-compact conformal field theories see 
\cite{Schomerus}.}. In the case of the free theory these requirements are
associated to the normalizability and unitarity of the string states. At the
level of the interacting theory, additional properties are requested, like
the closure of the fusion rules, the factorization properties of $N$-point
functions, etc.

The $SL(2,\mathbb{R})_{k}$ WZW model is behind the description of string
theory in both the 2D black hole background (through the coset construction)
and in $AdS_{3}$ space. These two models are closely related indeed, but
still different. In the case of the black hole, the states of the spectrum
are labeled by the index $j$ of the $SL(2,\mathbb{R})$ representations with
the indices $m$ and $\overline{m}$ falling in the lattice%
\begin{equation}
m-\overline{m}=n,\quad m+\overline{m}=-k\omega  \label{u}
\end{equation}%
with $n$ and $\omega $ being integer numbers, and the conformal dimension of
the vertex operators is given by%
\begin{equation}
h=-\frac{j(j+1)}{k-2}+\frac{m^{2}}{k}.  \label{harriba}
\end{equation}%
On the other hand, $AdS_{3}$ string theory can be described in terms of the
WZW model on the product between the coset $SL(2,\mathbb{R})_{k}/U(1)$ and a
time-like free boson \cite{HW}, so that the worldsheet theory turns out to
be the product between the time and the euclidean black hole. This can be
realized by adding the contribution\footnote{%
Besides, one can represent string theory in $AdS_{3}$ space in terms of the
Wakimoto free field realization mentioned above. In terms of these fields
the $AdS_{3}$ metric reads $ds^{2}=k\left( d\phi ^{2}+e^{2\phi }d\gamma d%
\overline{\gamma }\right) $.} $-\frac{1}{4\pi }\int d^{2}z\ \partial T%
\overline{\partial }T$ to the action (\ref{S0}) and by supplementing the
vertex operators with a factor $e^{i\sqrt{\frac{2}{k}}(m+\frac{k}{2}\omega
)T}$ that carries the charge under the field $T$. Thus, the vertex operators
on $AdS_{3}$ have conformal dimension given by%
\begin{equation}
h=-\frac{j(j+1)}{k-2}-m\omega -\frac{k}{4}\omega ^{2},  \label{h}
\end{equation}%
which corresponds to adding the conformal dimension $\delta h=-\frac{%
(m+k\omega /2)^{2}}{k}$ of the time-like part to the coset contribution (\ref%
{harriba}). In some sense, the string theory in the 2D black hole can be
thought of as having constrained the states of the theory in $AdS_{3}$ to
have vanishing bulk energy, $m+\overline{m}+k\omega =0$. In this way, one
has the theory on the background $time\times SL(2,\mathbb{R})_{k}/U(1)$\ as
an appropriate realization of sting theory in $AdS_{3}$ space \cite%
{MO1,GN2,GN3,GL}. However, before going deeper into the string
interpretation of the WZW model, some obstacles have to be overcame. In
fact, even in the case of the free string theory, the fact of considering
non-compact Lorentzian curved backgrounds is not trivial at all. The main
obstacle in constructing the space of states is the fact that, unlike what
happens in flat space, in curved space the Virasoro constraints are not
enough to decouple the negative-norm string states. In the early attempts
for constructing a consistent string theory in $AdS_{3}$, additional \textit{%
ad hoc} constraints were imposed on the vectors of the $SL(2,\mathbb{R})_{k}$
representations in order to decouple the ghosts. The vectors of $SL(2,%
\mathbb{R})$ representations are labeled by a pair of indices $j$ and $m$,
and thus such additional constraints (demanded as sufficient conditions for
unitarity) imply an upper bound for the index $j$ of certain
representations, and consequently an unnatural upper bound for the mass
spectrum. The modern approaches to the \textquotedblleft negative norm
states problem\textquotedblright\ also include such a kind of constraint on $%
j$, although this fact does not imply a bound on the mass spectrum as in the
old versions it did \cite{MO1}. The upper bound for the index $j$ of
discrete representations, often called \textquotedblleft unitarity
bound\textquotedblright , reads $1-k<2j<-1.$ In the case of Euclidean $%
AdS_{3}$, the spectrum of string theory is just given by the continuous
series of $SL(2,\mathbb{C})$, parameterized by the values $j=-\frac{1}{2}%
+i\lambda $ with $\lambda \in \mathbb{R}$ and by real $m$. On its turn, the
case of string theory in Lorentzian $AdS_{3}$ is richer and its spectrum is
composed by states belonging to both continuous $\mathcal{C}_{\lambda
}^{\alpha ,\omega }$ and discrete $\mathcal{D}_{j}^{\omega ,\pm }$ series.
The continuous series $\mathcal{C}_{\lambda }^{\alpha ,\omega }$ have states
with $j=-\frac{1}{2}+i\lambda $ with $\lambda \in \mathbb{R}$ and $m-\alpha
\in \mathbb{Z}$, with $\alpha \in \lbrack 0,1)\in \mathbb{R}$ (as in $SL(2,%
\mathbb{C})$, obviously). On the other hand, the states of discrete
representations $\mathcal{D^{\pm ,\omega }}_{j}$ satisfy $j=\pm m-n$ with $%
n\in \mathbb{Z}_{\geq 0}$. Other important ingredient for constructing the
Hilbert space is the index $\omega $ labeling the operators $\Phi _{j,m,\bar{%
m}}^{\omega }$. In the black hole background, $\omega $ turns out to be
given by (\ref{u}). In $AdS_{3}$, the quantum number $\omega $ is
independent of the bulk kinetic energy $m+\overline{m}$ and the bulk angular
momentum $m-\overline{m},$ contributing to the total energy as $m+\overline{m%
}+k\omega $. Then, the question arises as to how the index $\omega $ appears
in the Hilbert space of the $SL(2,\mathbb{R})_{k}$ WZW theory. The answer is
that in order to fully parameterize the spectrum in $AdS_{3}$ we have to
introduce the \textquotedblleft flowed\textquotedblright\ operators $\tilde{J%
}_{n}^{a}$ (with $a=3,-,+$) which are defined through the spectral flow
automorphism \cite{MO1} 
\begin{equation}
J_{n}^{3}\rightarrow \tilde{J}_{n}^{3}=J_{n}^{3}-\frac{k}{2}\omega \delta
_{n,0},\quad J_{n}^{\pm }\rightarrow \tilde{J}_{n}^{\pm }=J_{n\pm \omega
}^{\pm }  \label{arrova}
\end{equation}%
acting of the original $\hat{sl(2)}_{k}$ generators $J_{n}^{a}$, which
satisfy the Lie product that define the affine algebra%
\begin{equation}
\lbrack J_{n}^{-},J_{m}^{+}]=-2J_{n+m}^{3}+nk\delta _{n,-m},\qquad \lbrack
J_{n}^{3},J_{m}^{\pm }]=\pm J_{n+m}^{\pm },\qquad \lbrack
J_{n}^{3},J_{m}^{3}]=-n\frac{k}{2}\delta _{n,-m}.  \label{elalgebra}
\end{equation}%
Then, states $\left| \Phi _{j,m,\bar{m}}^{\omega }\right\rangle $ belonging
to the \textit{flowed} discrete representations $\mathcal{D}_{j}^{\pm
,\omega }$ are those obeying\footnote{%
or analogous relations for the Weyl reflected representations, namely $%
j\rightarrow -1-j$.} 
\begin{equation}
\tilde{J}_{0}^{\pm }\left| \Phi _{j,m,\bar{m}}^{\omega }\right\rangle =(\pm
j-m)\left| \Phi _{j,m\pm 1,\bar{m}}^{\omega }\right\rangle ,\quad \tilde{J}%
_{0}^{3}\left| \Phi _{j,m,\bar{m}}^{\omega }\right\rangle =m\left| \Phi
_{j,m,\bar{m}}^{\omega }\right\rangle  \label{t2}
\end{equation}%
and being annihilated by the positive modes, namely 
\begin{equation}
\tilde{J}_{n}^{a}\left| \Phi _{j,m,\bar{m}}^{\omega }\right\rangle =0\ ,\ \
\ n>0\ .  \label{t3}
\end{equation}%
States with $m=\pm j$ represent highest (resp. lowest) weight states, while
primary states of the continuous representations $\mathcal{C}_{\lambda
}^{\alpha ,\omega }$ are annihilated by all the positive modes. On the other
hand, the excited states in the spectrum are defined by acting with the
negative modes $J_{-n}^{a}$ ($n\in \mathbb{Z}_{>0}$) on the Kac-Moody
primaries $\left| \Phi _{j,m,\bar{m}}^{\omega }\right\rangle $; these
negative modes play the role of creation operators (\textit{i.e.} creating
the string excitation). The \textquotedblleft flowed
states\textquotedblright\ (namely those being primary vectors with respect
to the $\tilde{J}_{n}^{a}$ defined with $|\omega |>1$) are not primary with
respect to the $\hat{sl(2)}_{k}$ algebra generated by $J_{n}^{a}$, and this
is clear from (\ref{arrova}). However, highest weight states in the series $%
\mathcal{D}_{j}^{+,\omega }$ are identified with lowest weight states of $%
\mathcal{D}_{-k/2-j}^{-,\omega }$, which means that spectral flow with $%
|\omega |=1$ is closed among certain subset of Kac-Moody primaries.

The states belonging to discrete representations have a discrete energy
spectrum and represent the quantum version of those string states that are
confined in the centre of $AdS_{3}$ space; these are called
\textquotedblleft short strings\textquotedblright\ and are the counterpart
of those states that are confined close to the tip of the cigar geometry. On
the other hand, the states of the continuous representations describe
massive \textquotedblleft long strings\textquotedblright\ that can escape to
the infinity, where the theory is weakly coupled. In the case of the 2D
black hole, the index $\omega $ of these long strings has a clear
interpretation as an \textquotedblleft asymptotically topological" degree of
freedom (is not a topological one though). Because of the euclidean black
hole has the geometry of a semi-infinite cigar and thus looks like a
cylinder very far from the tip, the states in the asymptotic region have a
winding number around such cylinder. However, this is not strictly a
cylinder but has topology $\mathbb{R}^{2}$ instead of $\mathbb{R}\times
S^{1} $, so that, as it happens in $AdS_{3}$, the winding number
conservation can be in principle violated. Of course, this feasibility of
violating $\omega $ is not evident from the background (\ref{S0})-(\ref%
{graviton}), which is reliable only far from the tip of the cigar, but the
phenomenon can occur when string interactions take place. Instead, in the
sine-Liouville theory, the violation of the winding number is understood in
a clear way, as due to the explicit dependence on the T-dual direction $%
\tilde{X}$. We will return to this point later. Now, let us discuss the
string interactions in the black hole geometry.

\subsubsection{String amplitudes and correlation functions in the $SL(2,%
\mathbb{R})_{k}$ WZW theory}

The string scattering amplitudes in the 2D black hole background are given
by (the integration over the inserting points of) correlation functions in
the $SL(2,\mathbb{R})_{k}$ WZW theory. The first exact computation of such
WZW three and two-point functions was performed by K. Becker and M. Becker
in Refs. \cite{B,BB2}, and it was subsequently extended and studied in
detail in Refs. \cite{T1}-\cite{T3} by J. Teschner. The interaction
processes of winding string states were studied later in \cite{MO3,GN3},
after J. Maldacena and H. Ooguri proposed the inclusion of spectral flowed
states in the spectrum of the theory \cite{MO1}. Moreover, several
formalisms were employed to study the correlators in this non-compact CFT 
\cite{ADS3}-\cite{ADS3ultimo}. One of the most fruitful tools to work out
the functional form of these WZW correlators was the analogy between these
and Liouville correlators \cite{FZ,Teschner,Ponsot,ADS3}. Another useful
approach to compute the exact correlation functions is the free field
representation \cite{B,BB2,HOS,IOS,GN3,GN2,GL}, which for the WZW model
turns out to be similar to what we discussed for the Liouville theory. Let
us briefly describe how this \textquotedblleft free field computation" works
for the case of the two-point function: Consider the correlation functions
of exponential operators $\Phi _{j,m,\overline{m}}^{\omega }=e^{\sqrt{\frac{2%
}{k-2}}\widehat{j}\varphi -i\sqrt{\frac{2}{k}}mX-i\sqrt{\frac{2}{k}}(m+\frac{%
k}{2}\omega )T}$ (with $\widehat{j}=-1-j$) in the theory (\ref{S0})
perturbed by the operator (\ref{graviton}), namely%
\begin{equation}
\mathcal{O}=M\left( \sqrt{\frac{k-2}{2}}\partial \varphi +i\sqrt{\frac{k}{2}}%
\partial X\right) \left( \sqrt{\frac{k-2}{2}}\overline{\partial }\varphi +i%
\sqrt{\frac{k}{2}}\overline{\partial }X\right) e^{-\sqrt{\frac{2}{k-2}}%
\varphi }=M\ \beta \overline{\beta }e^{-\sqrt{\frac{2}{k-2}}\phi }.
\end{equation}%
Then, written in terms of the Wakimoto free fields\footnote{%
Please, do not mistake the Wakimoto field $\gamma $ (which is a local
function on the variable $z$) for the Euler $\gamma $-function introduced in
Eq. (\ref{introducidas}) (which is defined by $\gamma (x)=\Gamma (x)/\Gamma
(1-x)$). That is, the fields $\gamma $ in (\ref{FiRulete}) have to be
distinguished from the function $\gamma $ in (\ref{FiRulete2}). We preferred
to employ the standard notation here.} $\phi $, $\gamma $, and $\beta $,
such correlators read\footnote{%
In order to compare with the original computation in Ref. \cite{B} it is
necessary to consider the Weyl reflection $j\rightarrow -1-j$, which is a
symmetry of the formula for the conformal dimension, actually.} 
\begin{equation*}
\left\langle \Phi _{j,m,\overline{m}}^{\omega }(z_{1})\Phi _{j,-m,-\overline{%
m}}^{-\omega }(z_{2})\right\rangle _{WZW}=\Gamma (-s)\delta
(s+2j+1)\prod_{r=2}^{s}\int d^{2}\omega _{r}\left\langle \gamma
^{-1-j-m}(z_{1})\overline{\gamma }^{-1-j-\overline{m}}(z_{1})\right. \times
\end{equation*}%
\begin{eqnarray}
&&\times \gamma ^{-1-j+m}(z_{2})\overline{\gamma }^{-1-j+\overline{m}%
}(z_{2})\beta (w_{1})\overline{\beta }(w_{1})\prod_{r=2}^{s}\left. \beta
(w_{r})\overline{\beta }(w_{r})\right\rangle \times  \notag \\
&&\times \left\langle e^{-\sqrt{\frac{2}{k-2}}(j+1)\phi (z_{1})}e^{-\sqrt{%
\frac{2}{k-2}}(j+1)\phi (z_{2})}e^{-\sqrt{\frac{2}{k-2}}\phi (w_{1})}\right.
\prod_{r=2}^{s}\left. e^{-\sqrt{\frac{2}{k-2}}\phi (w_{r})}\right\rangle ,
\label{FiRulete}
\end{eqnarray}%
where the screening inserted at $w_{1}$ is then taken to be fixed at
infinity $w_{1}\rightarrow \infty $, while $z_{1}=0$ and $z_{2}=1$ as usual
(this is analogous to what we did when discussed the case of Liouville
partition function). It is easy to see that this can be solved by using the
(analytic extension of) Dotsenko-Fateev integrals, and one eventually finds%
\footnote{%
For instance, compare with formula (49) in Ref. \cite{GN3}, after the Weyl
reflection.} 
\begin{equation*}
\left\langle \Phi _{j,m,\overline{m}}^{\omega }(0)\Phi _{j,-m,-\overline{m}%
}^{-\omega }(1)\right\rangle _{WZW}=-\frac{\Gamma (-j-m)\Gamma (-j+m)}{%
\Gamma (j+1+\overline{m})\Gamma (j+1-\overline{m})}\times .
\end{equation*}%
\begin{equation}
\times \left( -\pi M\gamma \left( \frac{1}{k-2}\right) \right) ^{-1-2j}\frac{%
\gamma (2j+2)}{k-2}\gamma \left( \frac{2j+1}{k-2}\right) ,  \label{FiRulete2}
\end{equation}%
where the $m$-dependent $\Gamma $-functions stand from the combinatorial
problem of counting the different ways of (Wick) contracting the $\gamma $%
-functions with the $\beta $-functions in (\ref{FiRulete}). Expression (\ref%
{FiRulete2}) is the so called $SL(2,\mathbb{R})_{k}$ WZW reflection
coefficient $\mathcal{R}_{k}(j,m)$ and corresponds to the exact results for
the two-point function. Notice that, in particular, (\ref{FiRulete2})
contains the factor $\gamma \left( \frac{2j+1}{k-2}\right) $ that keeps
track of finite-$k$ effects. Analogously, the expression of the three-point
functions $\left\langle \Phi _{j_{1},m_{1},\overline{m}_{1}}^{\omega
_{1}}(z_{1})\Phi _{j_{2},m_{2},\overline{m}_{2}}^{\omega _{2}}(z_{2})\Phi
_{j_{3},m_{3},\overline{m}_{3}}^{\omega _{3}}(z_{3})\right\rangle _{WZW}$
can be found by these means \cite{BB2}.

Also, some features of the four-point function are known, as the physical
interpretation of its divergences \cite{MO3}, and the crossing symmetry \cite%
{Teschner}. In fact, our understanding of correlation functions in both the
2D black hole and $AdS_{3}$ backgrounds has substantially increased
recently, and we have a relatively satisfactory understanding of these
observables. Nevertheless, some features remain still open questions: One
puzzle is the factorization properties of the generic four-point function
and the closure of the operator product expansion of unitary states.
Addressing these questions would require a deeper understanding of the
analytic structure of the four-point function. The general expression for
the $N$-point functions for $N>3$ is not known; however, a new insight about
its functional form appeared recently due to the discovery of a new relation
between these and analogous correlators in Liouville field theory \cite%
{S,RT,R}. This relation between WZW and Liouville correlators is one of the
key points for what we are going to study in this paper. Let us give some
details about it.

\subsection{A connection between Liouville and WZW correlation functions}

Let us comment on the particular connection that exists between the
correlation functions of the two conformal theories we discussed above;
namely, between Liouville and $SL(2,\mathbb{R})_{k}$ WZW correlation
functions. This relation is a result recently obtained by S. Ribault and J.
Teschner, who have found a direct way of connecting correlators in both $%
SL(2,\mathbb{C})_{k}/SU(2)$ WZW and Liouville conformal theories \cite{RT,R}%
. The formula they proved is an improved version of a previous result
obtained by A. Stoyanovsky some years ago \cite{S}. The Ribault-Teschner
formula (whose more general form was presented by Ribault in Ref. \cite{R})
connects the $N$-point tree-level scattering amplitudes in Euclidean $%
AdS_{3} $ string theory to certain subset of $N+M$-point functions in
Liouville field theory, where the relation between $N$ and $M$ is determined
by the winding number of the interacting strings. Even though this formula
was proven for the case of the Euclidean target space, it is likely that an
analytic continuation of it also holds for the Lorentzian model. The
Ribault-Teschner formula reads as follows: If $\Phi _{j,m,\bar{m}}^{\omega }$
represent the vertex operators in the WZW model, and $V_{\alpha }$ represent
the vertex operators of Liouville theory, then it turns out that 
\begin{eqnarray}
\langle \prod_{i=1}^{N}\Phi _{j_{i},m_{i},\bar{m}_{i}}(z_{i})\rangle _{WZW}
&=&N_{k}(j_{1},...j_{N};m_{1},...m_{N})\prod_{r=1}^{M}\int d^{2}w_{r}\
F_{k}(z_{1},...z_{N};w_{1},...w_{M})\times  \notag \\
&&\times \langle \prod_{t=1}^{N}V_{\alpha _{t}}(z_{t})\prod_{r=1}^{M}V_{-%
\frac{1}{2b}}(w_{r})\rangle _{S_{L}[\mu ]}\ ,  \label{rrtt}
\end{eqnarray}%
with the normalization factor given by 
\begin{equation}
N_{k}(j_{1},...j_{N};m_{1},...m_{N})=\frac{2\pi ^{3-2N}b}{M!\ c_{k}^{M+2}}%
(\pi ^{2}\mu b^{-2})^{-s}\prod_{i=1}^{N}\frac{c_{k}\ \Gamma (-m_{i}-j_{i})}{%
\Gamma (1+j_{i}+\bar{m}_{i})}  \label{N}
\end{equation}%
and the $z$-dependent function given by 
\begin{eqnarray}
F_{k}(z_{1},...z_{N};w_{1},...w_{M}) &=&\frac{\prod_{1\leq
r<l}^{N}|z_{r}-z_{l}|^{k-2(m_{r}+m_{l}+\omega _{r}\omega _{l}k/2+\omega
_{l}m_{r}+\omega _{r}m_{l})}}{\prod_{1<r<l}^{M}|w_{r}-w_{l}|^{-k}%
\prod_{t=1}^{N}\prod_{r=1}^{M}|w_{r}-z_{t}|^{k-2m_{t}}}\times  \notag \\
&&\times \frac{\prod_{1\leq r<l}^{N}(\bar{z}_{r}-\bar{z}_{l})^{m_{r}+m_{l}-%
\bar{m}_{r}-\bar{m}_{l}+\omega _{l}(m_{r}-\bar{m}_{r})+\omega _{r}(m_{l}-%
\bar{m}_{l})}}{\prod_{1<r<l}^{M}(\bar{w}_{r}-\bar{z}_{t})^{m_{t}-\bar{m}_{t}}%
},  \label{FF}
\end{eqnarray}%
and where the parameter $b$ of the Liouville theory is related to the
Kac-Moody level $k$ through $b^{-2}=k-2$. The quantum numbers of the states
of both conformal models are related ones to each others through the simple
relation $\alpha _{i}=bj_{i}+b+b^{-2}/2\ ,\ $with$\ i=1,2,...N.$ The factor $%
c_{k}$ in (\ref{N}) is a $k$-dependent ($j$-independent) normalization; see 
\cite{R}. Furthermore, the following constraints also hold $m_{1}+...m_{N}=%
\bar{m}_{1}+...\bar{m}_{N}=\frac{k}{2}(N-M-2),$ $\omega _{1}+..\omega
_{N}=M+2-N$, $s=-b^{-1}(\alpha _{1}+...\alpha _{N})+b^{-2}\frac{M}{2}%
+1+b^{-2},$ where $s$ refers to the amount of screening operators $V_{b}=\mu
e^{\sqrt{2}b\varphi }$ to be included in the Liouville correlators in order
to get a non vanishing result, as in (\ref{pupo2}). Also notice that the
Liouville correlator in the r.h.s. of (\ref{rrtt}) contains $M$ degenerate
fields $V_{-1/2b}$ (i.e. states that contain null descendents in the
modulo), which have conformal dimension strictly lower than zero for
positive $b$. So that the formula (\ref{rrtt}) relates $N$-point functions
in the WZW theory to $M+N$-point functions in Liouville field theory.
Applications of (\ref{rrtt}) were discussed in \cite%
{Apl1,Apl2,YoYu,Takayanagi}, and ulterior generalizations were presented in 
\cite{disc1,disc2,HikidaSchomerus1}. The way of proving (\ref{rrtt}) was
making use of the relation existing between solutions to the BPZ\
differential equations (satisfied by the Liouville correlation functions
involved in (\ref{rt}), \cite{BPZ}) and the generalized KZ differential
equation (satisfied by the WZW correlators \cite{KZ,R}). This remarkable
trick allowed to demonstrate the map between correlators in both theories
even though one does not know the generic form of such observables in any of
the two cases.

The dictionary given by the formula (\ref{rrtt}) will play a crucial role in
proving the correspondence between the 2D black hole and the flat tachyonic
background we are interested in. Conversely, our result can be seen as a
free field realization of the Ribault-Teschner formula (\ref{rrtt}). In
fact, in section 4 we will describe how (\ref{rrtt}) can be thought of as an
identity between the $SL(2,\mathbb{R})_{k}$ WZW theory and a CFT of the form 
$Liouville\times U(1)\times \mathbb{R}$, for which the $U(1)$ dependences of
the correlators factorize out yielding the piece $%
F_{k}(z_{1},...z_{N};w_{1},...w_{M})$ in (\ref{FF}). In this realization,
the operators $V_{-1/2b}$ are seen as $M$ additional screening currents. The
details of this can be found in subsection 4.3. First, let us discuss the
FZZ duality.

\section{The FZZ dual for the 2D black hole}

In this section we will study the (standard) FZZ dual for the
two-dimensional black hole, namely the sine-Liouville field theory. In
particular, we will discuss it as an example of tachyon background for the
2D string theory.

\subsection{Tachyon-like backgrounds in 2D string theory}

Let us consider the non-linear $\sigma $-model on a generic curved target
space of metric $G_{\mu \nu },$ and in presence of both dilatonic $\Phi $
and tachyonic $\mathcal{T}$ backgrounds\footnote{%
As it is known, in two dimensions the expression \textquotedblleft
tachyonic" has to be understood just formally, since the tachyon is massless
in $D=2$; see \cite{fin} for an illustrative example.}. If we supplement the
worldsheet action (\ref{Polyakov2}) with the tachyonic term, the $\sigma $%
-model takes the form%
\begin{equation}
S_{P}=\frac{1}{4\pi }\int d^{2}z\left( G_{\mu \nu }(X)\partial X^{\mu }%
\overline{\partial }X^{\nu }+R\Phi (X)+\mathcal{T}(X)\right) ,
\label{Polyakov}
\end{equation}%
where, as before, $\mu ,\nu =\{1,2\};$ and where we adopt the convention $%
X^{1}\equiv X$ and $X^{2}\equiv \varphi $ representing the two coordinates
that parameterize the target space. Thus, conformal invariance at quantum
level demands the vanishing of the $\beta $-functions for the action (\ref%
{Polyakov}); and for the tachyon field, the one-loop linearized $\beta $%
-function reads \cite{Polchinski}%
\begin{equation}
\beta ^{\mathcal{T}}=-\nabla _{\mu }\nabla ^{\mu }\mathcal{T}+2\nabla _{\mu
}\Phi \nabla ^{\mu }\mathcal{T}-2\mathcal{T}=0,  \label{labeta}
\end{equation}%
where higher powers of $\mathcal{T}$ were neglected. This equation, together
with the one-loop $\beta $-functions for the metric and the dilaton, admit
solutions of the form%
\begin{equation}
G_{\mu \nu }=\delta _{\mu \nu },\qquad \Phi (\varphi )=\frac{Q}{\sqrt{2}}%
\varphi ,\qquad \mathcal{T}(X,\varphi )=\sum_{n}\lambda _{n}\,e^{\sqrt{2}%
a_{n}\varphi +i\sqrt{2}b_{n}X},  \label{solutione}
\end{equation}%
with $a_{n}(Q-a_{n})+b_{n}^{2}=1$, $Q=2$. Here, coefficients $\lambda _{n}$
are real numbers that can be regarded as the Fourier modes of the tachyon
potential. The tachyon momenta $b_{n}$ are chosen to be consistent with the
compactification conditions for the $X$ direction; in particular, here we
will consider $b_{n}=n\sqrt{k}/2$, and the tachyon potential will be of the
Toda-like form%
\begin{equation}
\mathcal{T}(X,\varphi )=\sum_{n=-\infty }^{\infty }\lambda _{n}e^{\sqrt{2}%
(1\pm \sqrt{k}|n|/2)\varphi +i\sqrt{k/2}nX};  \label{torbellino2}
\end{equation}%
see (\ref{cos}) and (\ref{O}) below. Background (\ref{solutione}) is the
type of configuration we will deal with. A particular case of interest is
the sine-Liouville theory, which we discuss below.

\subsection{Sine-Liouville theory and the FZZ conjecture}

\subsubsection{Sine-Liouville field theory}

Sine-Liouville theory is a particular case of tachyon-like background, and,
according to the FZZ conjecture, this is dual to the 2D string theory on the
black hole spacetime. Sine-Liouville theory corresponds to perturb the free
action $S_{0}$ with the operator%
\begin{equation}
\mathcal{O}_{\widetilde{\lambda }_{-1}=\widetilde{\lambda }_{+1}=\lambda
}=4\lambda e^{-\sqrt{\frac{k-2}{2}}\varphi }\cos \left( \sqrt{k/2}\widetilde{%
X}\right) ,  \label{cos}
\end{equation}%
which is convenient to write as%
\begin{equation}
\mathcal{O}_{\widetilde{\lambda }_{-1}=\widetilde{\lambda }_{+1}=\lambda
}=2\lambda e^{-\sqrt{\frac{k-2}{2}}\varphi +i\sqrt{\frac{k}{2}}\widetilde{X}%
}+2\lambda e^{-\sqrt{\frac{k-2}{2}}\varphi -i\sqrt{\frac{k}{2}}\widetilde{X}%
},  \label{cos2}
\end{equation}%
where $\widetilde{X}=X_{L}(z)-X_{R}(\overline{z})$. The interaction term (%
\ref{cos}) resembles both the sine-Gordon and the Liouville field theories,
and this is the reason of the name of \textquotedblleft sine-Liouville".
Actually, this theory corresponds to the sine-Gordon model coupled to
two-dimensional gravity.

The sine-Liouville interaction (\ref{cos}) can be thought of as a particular
case of the action (\ref{S}) if the $X$ field is replace by its T-dual $%
\widetilde{X}$. It would correspond to the coupling $\widetilde{\lambda }%
_{n} $ $=$ $\lambda \left( \delta _{n+1}+\delta _{n-1}\right) $. This field
theory describes the phase of vortex condensation in the 2D string theory.
Unlike the euclidean black hole geometry, whose topology is $\mathbb{R}^{2}$%
, sine-Liouville theory is an interacting CFT formulated on the topology $%
\mathbb{R}\times S^{1}$. The distinct topologies arise because the angular
direction of the (simple connected) cigar is the one of the duality
transformation. Notice also that sine-Liouville interaction term is not
bounded from below, and this is ultimately related to the $\mathbb{R}^{2}$
topology of the cigar too.

Sine-Liouville theory and its relation to the 2D black hole have been
extensively studied in the last six years, and, as we mentioned, this has
led to the formulation of the matrix model for the black hole \cite{KKK}.
The matrix model turned out to be a very important tool for studying black
hole physics in string theory; in particular, it permitted to address the
question about the black hole formation in string theory \cite%
{MaldacenaStrominger}. Matrix model formulation also enabled to study the
integrability of the theory from a different point of view\footnote{%
The black hole turns out to be dual to the perturbed $c=1$ theory, and, on
the other hand, the $c=1$ theory perturbed by the vortex or tachyon
potential turns out to be integrable with the integrable structure described
in terms of the Toda hierarchy \cite{KKK}.}, and we emphasize that all this
was possible because of FZZ duality.

\subsubsection{The Fateev-Zamolodchikov-Zamolodchikov conjecture}

FZZ duality is a strong-weak duality. The semiclassical limit of
sine-Liouville theory corresponds to the limit $k\rightarrow 2,$ where the
black hole is highly curved. Conversely, the semiclassical limit of the
black hole theory corresponds to the large $k$ regime where the
sine-Liouville wave function is strongly suppressed in the $\varphi $
direction. Perhaps, the correct way of thinking FZZ duality is that the full
theory is actually described by both the WZW and sine-Liouville models, and
each of them dominates the dynamics of the theory in a different regime
(where the corresponding action is reliable as a good approximation).
However, it is worth mentioning that both theories have control on the
observables beyond the regime in which one would naively expect so. For
instance, even though one would expect the black hole $\sigma $-model action
to describe the theory only in the large $k$ regime, it turns out that the
Coulomb gas computation of correlation function using the screening operator
behaving like $\sim e^{-\sqrt{\frac{2}{k-2}}\varphi }$ do reproduce the
exact result, including finite-$k$ effects\footnote{%
We have exemplified this in the previous section by computing the two-point
function.} \cite{B,BB2,GN3}. Besides, the same feature occurs for the
computation in sine-Liouville theory \cite{FH}. This sourprising feature is
due to the analytic extension of the Coulomb gas type expressions, which is
powerful enough to reconstruct the exact expressions of the correlators.
This is precisely what permitted to perform consistency checks of the
conjecture. The interplay between perturbative poles and $k$-dependent poles
in correlation functions of both models was first discussed in \cite{KKK},
where it was shown that the poles of bulk amplitudes in sine-Liouville
precisely reproduce non-perturbative (finite-$k$ effects) poles of WZW
correlators\footnote{%
However, again, it is important to emphasize that such finite-$k$ poles can
be directly obtained by considering the perturbative action of the WZW
theory \cite{B}. For instance, in Ref. \cite{GN3} it was shown that the
computation in the WZW model involving operators behaving like $\sim e^{-%
\sqrt{2(k-2)}\varphi }$ exactly agree with those originally computed in \cite%
{B}, even though the dependence on $k$ is the opposite to the one appearing
in (\ref{graviton}).}.

Strong-weak FZZ correspondence turns out to be a very important piece for
our understanding of black hole physics in string theory. So, let us briefly
discuss how such correspondence works operatively. First, we present the
main ingredients: The sine-Liouville vertex operators we have to consider
are those of the form\footnote{%
Here, we are not explicitly writing the antiholomorphic contribution $e^{i%
\sqrt{2/k}\bar{m}X}$ for short; it has to be understood in all the formulae
below. Besides, let us notice that vertex (\ref{caritaredonda}) would
receive an extra piece $e^{i\sqrt{\frac{2}{k}}(m+k\omega /2)T+i\sqrt{\frac{2%
}{k}}(\bar{m}+k\omega /2)T}$ in the case that the theory one considers is
the product between the sine-Liouville action and the time direction $T$.}%
\begin{equation}
\mathcal{T}_{j,m,\overline{m}}=e^{\sqrt{\frac{2}{k-2}}j\varphi +i\sqrt{\frac{%
2}{k}}mX}.  \label{caritaredonda}
\end{equation}%
The spectrum of the theory contains states obeying $m-\overline{m}=k\omega $
and $m+\overline{m}=n,$ with integers $n$ and $\omega $. Operators (\ref%
{caritaredonda}) have conformal dimension

\begin{equation*}
h=-\frac{j(j+1)}{k-2}+\frac{m^{2}}{k},
\end{equation*}%
and the coincidence with (\ref{harriba}) shows the convenience of this
notation. A crucial observation is that the sine-Liouville theory presents
symmetry under the $\widehat{sl}(2)_{k}$ affine algebra, and this can be
realized by free field techniques by defining \cite{Satohviejo}%
\begin{equation}
J^{\pm }(z)=\left( -i\sqrt{\frac{k}{2}}\partial X\pm \sqrt{\frac{k-2}{2}}%
\partial \varphi \right) e^{\mp i\sqrt{\frac{2}{k}}(T+X)},\qquad J^{3}(z)=i%
\sqrt{\frac{k}{2}}\partial T.  \label{currents}
\end{equation}%
These currents satisfy the OPE 
\begin{equation*}
J^{3}(z)J^{\pm }(w)=\pm \frac{1}{(z-w)}J^{\pm }(w)+...,\qquad
J^{3}(z)J^{3}(w)=-\frac{k/2}{(z-w)^{2}}+...,
\end{equation*}%
\begin{equation*}
J^{-}(z)J^{+}(w)=\frac{k}{(z-w)^{2}}-\frac{2}{(z-w)}J^{3}(w)+...,
\end{equation*}%
and thus realize (\ref{elalgebra}) by means of the relation $J_{n}^{a}=\frac{%
1}{2\pi i}\oint dz\ z^{-1-n}J^{a}(z)$. It is possible to verify that
sine-Liouville interaction commutes with these currents, namely that the
OPEs yield regular terms. This matching of symmetries is an important
necessary condition for the equivalence to the WZW theory. The next step
would be that of proposing a dictionary between observables: According to
FZZ\ prescription, operators (\ref{caritaredonda}) are associated in
one-to-one correspondence to those operators that expand $SL(2,\mathbb{R}%
)_{k}$ representations in the theory on the coset \cite{KKK}, namely%
\begin{equation*}
\mathcal{T}_{j,m,\overline{m}}\leftrightarrow \Phi _{j,m,\overline{m}},
\end{equation*}%
where $\Phi _{j,m,\overline{m}}$ are the vertex operators on the coset
theory $SL(2,\mathbb{R})_{k}/U(1)$, defined through their relation to the $%
SL(2,\mathbb{R})_{k}$ vertex, namely $\Phi _{j,m,\overline{m}}^{\omega
}=\Phi _{j,m,\overline{m}}\times e^{i\sqrt{\frac{2}{k}}(m+\frac{k}{2}\omega
)T}$. Then, once the sine-Liouville operators were introduced, one can
undertake the task of performing perturbative checks of the duality. To do
this, one should compare the analytic structure of correlation functions in
both conformal models; but, first, one has to know how to compute such
quantities. So, let us review the computation of correlators for the
sine-Liouville field theory.

\subsubsection{Correlation functions in sine-Liouville theory}

Correlation functions in the sine-Liouville theory are assumed to reproduce
the analytic structure of the WZW analogues. The former can be computed by
standard Coulomb gas techniques, and the precise prescription was studied in
Ref. \cite{FB,FH}. In the case $N\leq 3$ these correlators were explicitly
integrated. In general, $N$-point sine-Liouville amplitudes are expected to
exhibit poles at $s_{-}+s_{+}=\frac{2}{k-2}\left(
1+\sum_{i=1}^{N}j_{i}\right) $, where the residues turn out to be expressed
in terms of multiple integrals over the whole complex plane. These read

\begin{eqnarray}
A_{(j_{1},...j_{N}|z_{1},...z_{N})}^{sine-L} &=&\frac{\lambda ^{\frac{2}{k-2}%
(j_{1}+...j_{N}+1)}}{s_{-}!s_{+}!}\prod_{r=1}^{s_{+}}\int
d^{2}v_{r}\prod_{t=1}^{s_{-}}\int d^{2}v_{t}\left\langle \mathcal{T}%
_{j_{1},m_{1},\overline{m}_{1}}(z_{1})\mathcal{T}_{j_{2},m_{2},\overline{m}%
_{2}}(z_{2})...\right.  \notag \\
&&...\mathcal{T}_{j_{N},m_{N},\overline{m}_{N}}(z_{N})\prod_{r=1}^{s_{+}}%
\mathcal{T}_{1-\frac{k}{2},\frac{k}{2},-\frac{k}{2}}(u_{r})%
\prod_{t=1}^{s_{-}}\left. \mathcal{T}_{1-\frac{k}{2},-\frac{k}{2},\frac{k}{2}%
}(v_{t})\right\rangle _{S_{[\lambda =0]}}  \label{UUU}
\end{eqnarray}%
with $S_{[\lambda =0]}=S_{0}$, yielding%
\begin{equation*}
A_{(j_{1},...j_{N}|z_{1},...z_{N})}^{sine-L}=\frac{\lambda ^{\frac{2}{k-2}%
(j_{1}+...j_{N}+1)}}{\Gamma (s_{-}+1)\Gamma (s_{+}+1)}\prod_{a<b}^{N-1,N}%
\left| z_{a}-z_{b}\right| ^{-\frac{4j_{a}j_{b}}{k-2}}\left(
z_{a}-z_{b}\right) ^{\frac{2}{k}m_{a}m_{b}}\left( \bar{z}_{a}-\bar{z}%
_{b}\right) ^{\frac{2}{k}\bar{m}_{a}\bar{m}_{b}}\times
\end{equation*}%
\begin{eqnarray}
&&\ \times \prod_{r=1}^{s_{+}}\int d^{2}u_{r}\prod_{l=1}^{s_{-}}\int
d^{2}v_{l}\prod_{r<t}^{s_{+}-1,s_{+}}\left| u_{r}-u_{t}\right|
^{2}\prod_{l<t}^{s_{-}-1,s_{-}}\left| v_{t}-v_{s}\right|
^{2}\prod_{l=1}^{s_{-}}\prod_{r=1}^{s_{+}}\left| v_{l}-u_{r}\right|
^{2-2k}\times  \notag \\
&&\ \times \prod_{a=1}^{N}\prod_{r=1}^{s_{+}}\left| z_{a}-u_{r}\right|
^{2(j_{a}+m_{a})}\left( \bar{z}_{a}-\bar{u}_{r}\right) ^{m_{a}-\overline{m}%
_{a}}\prod_{b=1}^{N}\prod_{l=1}^{s_{-}}\left| z_{b}-v_{l}\right|
^{2(j_{b}-m_{b})}\left( \bar{z}_{b}-\bar{v}_{l}\right) ^{m_{b}-\overline{m}%
_{b}},  \label{21}
\end{eqnarray}%
that follows from the Wick contraction of operators $\mathcal{T}_{j,m,%
\overline{m}}$ and $\mathcal{T}_{1-\frac{k}{2},\pm \frac{k}{2},\pm \frac{k}{2%
}}$. The poles that correspond to bulk amplitudes in sine-Liouville theory
can be shown to arise though the integration over the zero-mode of the field 
$\varphi $ \cite{diFK}. In the case $N=3$ the pole structure of (\ref{21})
was shown to agree with that of the black hole theory, for which the finite-$%
k$ poles represent non-perturbative worldsheet effects. This non-trivial
matching between analytic structures was one of the strongest evidence in
favor of the FZZ conjecture at perturbative level \cite{KKK,FH,GL}. An
important piece of information is encoded in the fact that the
sine-Liouville correlators scale as $\lambda ^{\frac{2}{k-2}%
(j_{1}+j_{2}+...j_{N}+1)}$ while the black hole correlators scale as $M^{1+%
\widehat{j}_{1}+\widehat{j}_{2}+...\widehat{j}_{N}}$. In particular, it
tells us something about how the sine-Liouville correlators behave in the
large $k$ limit.

In Ref. \cite{FH} the authors translated the integrals $\prod_{r,l}\int
d^{2}u_{r}\int d^{2}v_{l}$ in (\ref{UUU}) into the product of contour
integrals. In this way, the integral representation above results described
by standard techniques developed in the context of rational conformal field
theory. Such techniques were used to evaluate the correlators to give a
formula for the contour integrals. The first step in the calculation is to
decompose the $u_{r}$ complex variables (resp. $v_{l}$) into two independent
real parameters (\textit{i.e.} the real and imaginary part of $u_{r}$) which
take values in the whole real line. Secondly, a Wick rotation for the
imaginary part of $(u_{r})$ has to be performed in order to introduce a
shifting parameter $\varepsilon $ which is ultimately used to elude the
poles in $z_{a}$. Then, the contours are taken in such a way that the poles
at $v_{r}\rightarrow z_{a}$ are avoided by considering the alternative order
with respect to this inserting points. The details of the prescription can
be found in the section 3 of Ref. \cite{FH}; see also Ref. \cite{FH2}.

\subsubsection{On the violation of the winding number conservation}

Now, let us return to the feature of the violation of winding conservation.
From the point of view of the sine-Liouville field theory the violation of
the total winding number in (\ref{UUU}) is given by $\sum_{a=1}^{N=3}\omega
_{a}=k^{-1}\sum_{a=1}^{N=3}(m_{a}+\overline{m}_{a})=s_{-}-s_{+}$ and comes
from the insertion of a different amount of screening operators $s_{-\text{ }%
}$ and $s_{+}$. It can be proven that for the three-point functions, the
winding can be violated up to $|\sum_{a=1}^{N=3}\omega _{a}|\leq N-2=1$ and,
presumably, this is the same for generic $N$. The key point for obtaining
such a constraint is noticing that the integrand that arises in the Coulomb
gas-like prescription contains contributions of the form%
\begin{equation*}
\int d^{2}v_{r}d^{2}v_{t}|v_{r}-v_{t}|^{2}...
\end{equation*}%
that come from the product expansion of two operators $\mathcal{T}_{1-\frac{k%
}{2},\pm \frac{k}{2},\pm \frac{k}{2}}$ inserted at the points $v_{r}$ and $%
v_{t}$ for $0\leq r,t\leq s_{-}$ (and the same for the points $u_{l}$ with $%
0\leq l\leq s_{+}$), and where the dots \textquotedblleft $...$%
\textquotedblright\ stand for \textquotedblleft other
dependences\textquotedblright\ on $v_{r}$ and $u_{l}$. As explained in \cite%
{FH}, the integral vanishes for certain alignments of contours\ due to the
fact that the exponent of $|v_{r}-v_{t}|$ is $+2$. Conversely, in the case
where such exponent is generic enough (let us say $2\rho $, following the
notation of \cite{FH}), the integral has a phase ambiguity due to the
multi-valuedness of $|v_{r}-v_{t}|^{2\rho }$ in the integrand. Then, those
integrals containing two contours of $v_{r}$ and $v_{t}$ just next to each
other vanish, and this precisely happens for all the contributions of those
correlators satisfying $|s_{+}-s_{-}|=|\sum_{a=1}^{N}\omega _{a}|>N-2$. This
led Fukuda and Hosomichi to prove that, for the three-point function, there
are only three terms that contribute: one with $\sum_{a=1}^{3}\omega _{a}=1$%
, a second with $\sum_{a=1}^{3}\omega _{a}=-1,$ and the conserving one, $%
\sum_{a=1}^{3}\omega _{a}=0$. A similar feature is exhibited in the
\textquotedblleft twisted" sine-Liouville model we will consider in the next
section, see \cite{0511252}.

On the other hand, one can wonder about how the violation of the winding
number conservation is seen from the point of view of the black hole theory,
where, unlike what happens in sine-Liouville theory, the action does not
seem to break the winding conservation. That it, even though the geometric
reason why the winding is not conserved in the cigar is quite clear, it is
not obvious how to understand such non-conservation in the calculation of
correlators. The answer to this puzzle was first given in Ref. \cite{FZZ},
and subsequently reviewed in \cite{MO3}. In fact, the computation of the
winding violating correlators in the WZW theory is far from being as simple
as in the case of sine-Liouville theory. In the WZW theory such computation
requires the insertion of one additional operator for each unit in which the
winding number is being violated. This additional operator is the often
called \textquotedblleft spectral flow operator" $\Phi _{-\frac{k}{2},\pm 
\frac{k}{2},\pm \frac{k}{2}}^{1}$, and this is an auxiliary operator that
plays the role of changing (in one unit) the winding number $\omega $ of a
given $SL(2,\mathbb{R})$-state involved in the correlator. The spectral flow
operator corresponds to a conjugate representation of the identity operator,
so it has conformal dimension zero. For instance, the three-point scattering
amplitudes (violating winding in one unit) in the 2D black hole would be
actually given in terms of a four-point correlation functions involving a
fourth dimension-zero operator $\Phi _{-\frac{k}{2},\pm \frac{k}{2},\pm 
\frac{k}{2}}^{1}$, after extracting the appropriate divergent factor coming
from the coincidence limit of spectral flow operator and the evaluation at $%
m=\bar{m}=\pm k/2;$ see \cite{MO3,YoYu} for the details.

Regarding the computation of correlation functions where the winding number
conservation is violated, let us mention that the most simple way of
computing such observables is that of making use of the twisted dual model
we will introduce in the next section. Perhaps this is the most useful
application of it, and we will comment on this feature later. Now, let us to
introduce the new dual model for the 2D black hole; which we will call the
\textquotedblleft twisted model" because it involves momentum modes of the
higher sector $n=2$.

\section{A twisted dual for the 2D black hole}

As said, we will now discuss an alternative dual description of string
theory in the 2D black hole ($\times $ $time$). First, we will introduce a
family of perturbations of the linear dilaton background (\ref{S0}) and, in
particular, we will introduce the perturbation that corresponds to the
twisted version of the sine-Liouville model which we want to relate to the
black hole $\sigma $-model. After doing this, we will make the precise
statement of such duality and show how to prove it by using the formula (\ref%
{rrtt}).

\subsection{Perturbations of higher winding and momentum modes}

\subsubsection{Momentum mode perturbations}

Let us begin by considering a rather general deformation of the theory (\ref%
{S0}), including higher modes of momentum and winding. The interaction term
in (\ref{S}) is given by the operator%
\begin{equation}
\mathcal{O}_{\lambda _{n}}=\sum_{n=-\infty }^{\infty }\lambda _{n}e^{-\frac{%
\alpha _{n}}{\sqrt{2}}\varphi +in\sqrt{\frac{k}{2}}X},  \label{O}
\end{equation}%
for which the condition $\lambda _{n}=\lambda _{-n}$ is required to be real.
Each term in this sum represents a marginal deformation of the linear
dilaton theory (\ref{S0}), and if the T-dual direction $\widetilde{X}$ is
considered instead of $X$ then this operator describes the sine-Liouville
field theory in the particular case $\widetilde{\lambda }_{n=1}=\widetilde{%
\lambda }_{n=-1}\neq 0$. The case $n=0$ is also included in the sum. In that
case the exponent is given by $\alpha _{n=0}=\frac{1+\sqrt{9-4k}}{\sqrt{k-2}}
$, so that it is real (represents a \textquotedblleft Liouville-like wall
potential") only for values $k\leq 9/4$. The value that saturates this
bound, $k=9/4,$ precisely corresponds to the black hole background, i.e. for
which the central charge of the coset $SL(2,\mathbb{R})_{k}/U(1)$ itself
turns out to be $26$. At $k=9/4$ the interaction term for $n=0$ turns out to
be $e^{-\sqrt{2}\varphi }$, i.e. the cosmological constant. We have to point
out that for $k=9/4$ the interaction (\ref{O}) agrees with the
two-dimensional string theory in an arbitrary winding background studied by
V. Kazakov, I. Kostov and D. Kutasov in\footnote{%
See formula (3.19) in Ref. \cite{KKK} and notice that the notation there
relates to the one employed here by $\varphi =\sqrt{2}\phi .$} Ref. \cite%
{KKK}. That is, for $k=9/4$ operator (\ref{O}) reads\footnote{%
Actually, the contribution $n=0$ at the point $k=9/4$ leads to the operator $%
\varphi e^{-\sqrt{2}\varphi }$ instead of $e^{-\sqrt{2}\varphi }$. This
comes from the fact that there are two possible values for $\alpha
_{n=0}=(1\pm \sqrt{9-4k})/\sqrt{k-2}$ which coincide (a resonance) in the
limit $k\rightarrow 9/4$ producing a degenerangy analogous to the case of
the Liouville cosmological term in the $b\rightarrow 1$ limit \cite{MTV}.
Also notice the difference between the signs of the exponents of (\ref%
{torbellino2}) and (\ref{torbellino1}); which is due to the sign of the
background charge in each case.}%
\begin{equation}
\mathcal{O}_{\lambda _{n}}=\lambda _{0}\varphi e^{-\sqrt{2}\varphi
}+\sum_{n\neq 0}\lambda _{n}e^{(|n|R-\sqrt{2})\varphi +inRX}.
\label{torbellino1}
\end{equation}%
with $R=\sqrt{k/2}=3/2\sqrt{2}$. The matrix model incorporating these
perturbations is constructed by implementing a deformed version of the Haar
measure on the $U(N)$ group manifold. The details of the matrix model
construction can be found in \cite{KKK}; here we will not discuss the
subject beyond the scope of the continuous limit.

\subsubsection{Adding vortex type perturbations}

It is instructive to explore other deformations. For instance, let us
consider the more general family%
\begin{equation}
\mathcal{O}_{\lambda _{n},\widetilde{\lambda }_{n}}=\sum_{n\neq 0}e^{-\frac{%
\alpha _{n}^{(-)}}{\sqrt{2}}\varphi }\left( \lambda _{n}^{(-)}e^{in\sqrt{%
\frac{k}{2}}X}+\widetilde{\lambda }_{n}^{(-)}e^{in\sqrt{\frac{k}{2}}%
\widetilde{X}}\right) +\sum_{n\neq 0}e^{-\frac{\alpha _{n}^{(+)}}{\sqrt{2}}%
\varphi }\left( \lambda _{n}^{(+)}e^{in\sqrt{\frac{k}{2}}X}+\widetilde{%
\lambda }_{n}^{(+)}e^{in\sqrt{\frac{k}{2}}\widetilde{X}}\right) ,
\label{pupo}
\end{equation}%
with $\alpha _{n}^{(\pm )}=\widehat{Q}(1\mp \sqrt{1+(kn^{2}-4)(k-2)}),$ so
that (\ref{O}) corresponds to the branch $\alpha _{n}^{(-)}$. In the 2D
black hole, couplings $\lambda _{n}$ turn the momentum modes on, while $%
\widetilde{\lambda }_{n}$ are the couplings of vortex operators turning
winding modes on, instead. Notice that perturbation (\ref{pupo})\ not only
includes the usual sine-Liouville interaction $\alpha _{\pm 1}^{(-)}$, but
also includes the dual sine-Liouville interaction introduced by A.
Mukherjee, S. Mukhi and A. Pakman in Ref. \cite{MMP} when the modes $\alpha
_{\pm 1}^{(+)}$ are considered\footnote{%
Notice that the notation in \cite{MMP} relates with ours here by $\varphi =-%
\sqrt{2}\phi .$}. Operators of the branches $\alpha _{n}^{(\pm )}$ have a
large $k$ behavior $\sim e^{\pm \sqrt{\frac{k}{2}}|n|\varphi }$, so that
only those of the branch $\alpha _{n}^{(-)}$ decrease for large $\varphi $
(where the theory is weakly coupled) in the black hole semiclassical limit $%
k\rightarrow \infty $. For our purpose, the interesting operators are those
having momentum $n=2$. In particular, here we are mainly interested in the
case $\alpha _{2}^{(-)}=\frac{2}{\sqrt{k-2}}(k-1)$; this is the one that
will enable us to present an alternative dual description for the 2D black
hole. For notational convenience, let us point out that operators with
momentum $\alpha _{n}^{(\pm )}$ can be written as 
\begin{equation*}
\mathcal{T}_{j_{n}^{\pm },m_{n},m_{n}}=e^{-\frac{\alpha _{n}^{(\pm )}}{\sqrt{%
2}}\varphi +in\sqrt{\frac{k}{2}}X},\quad \mathcal{T}_{j_{n}^{\pm
},m_{n},-m_{n}}=e^{-\frac{\alpha _{n}^{(\pm )}}{\sqrt{2}}\varphi +in\sqrt{%
\frac{k}{2}}\widetilde{X}},
\end{equation*}%
so carrying momentum $j_{n}^{\pm }=-\frac{1}{2}\pm \frac{1}{2}\sqrt{%
1+(k-2)(kn^{2}-4)}$ and momentum or winding number $m_{n}\pm \bar{m}_{n}=kn$%
. Here we will discuss how certain correlation functions of the model
defined by the action (\ref{S}), when the momentum modes $n=2$ (represented
by operators $\mathcal{T}_{1-k,k,k}$) and $n=1$ (respectively represented by 
$\mathcal{T}_{1-\frac{k}{2},\frac{k}{2},\frac{k}{2}}$) are turned on,
precisely agree with the correlation functions of the $SL(2,\mathbb{R})_{k}$
WZW model. So, now we are ready to make the main statement about the
correspondence and describe the precise prescription for computing the
correlators.

\subsection{Statement of the correspondence}

\subsubsection{Preliminary: some definitions}

We will discuss how the particular deformation of the family (\ref{O}) given
by $\lambda _{n}=\mu \delta _{n-2}+\lambda \delta _{n-1}$ is dual to the 2D
black hole, in a similar way as the sine-Liouville model is so. That is, by
\textquotedblleft dual" we mean that there exists a direct correspondence
between correlation functions of both CFTs at the level of the sphere
topology. Then, the interaction operator we will consider is 
\begin{equation}
\mathcal{O}_{\lambda _{1}=\lambda ,\lambda _{2}=\mu }=\lambda e^{-\sqrt{%
\frac{k-2}{2}}\varphi +i\sqrt{\frac{k}{2}}X}+\mu e^{-\sqrt{\frac{2}{k-2}}%
(k-1)\varphi +i\sqrt{2k}X},  \label{Uf}
\end{equation}%
where the scaling relation between the coupling constants $\mu $ and $%
\lambda $ goes like $\lambda ^{2}=a_{k}\mu $ with a $k$-dependent
proportionality factor $a_{k}$ that will be specified below. In the large $k$
limit this operator behaves like 
\begin{equation*}
\mathcal{O}_{\lambda _{1}=\lambda ,\lambda _{2}=\mu }\sim \lambda e^{-\sqrt{%
k/2}(\varphi -iX)}+\frac{1}{a_{k}}(\lambda e^{-\sqrt{k/2}(\varphi
-iX)})^{2}\sim \lambda e^{-\sqrt{k/2}(\varphi -iX)}+\frac{1}{a_{k}}\left( 
\mathcal{O}_{\lambda _{1}=\lambda ,\lambda _{2}=0}\right) ^{2}.
\end{equation*}%
Also notice that 
\begin{equation}
\mathcal{T}_{1-\frac{k}{2},\frac{k}{2},\frac{k}{2}}=e^{-\sqrt{\frac{k-2}{2}}%
\varphi +i\sqrt{\frac{k}{2}}X},\qquad \mathcal{T}_{1-k,k,k}=e^{-\sqrt{\frac{2%
}{k-2}}(k-1)\varphi +i\sqrt{2k}X},  \label{Considering}
\end{equation}%
so that 
\begin{equation}
{\mathcal{O}}_{\lambda _{1}=\lambda ,\lambda _{2}=\mu }=\lambda \mathcal{T}%
_{1-\frac{k}{2},\frac{k}{2},\frac{k}{2}}+\frac{\lambda ^{2}}{a_{k}}\mathcal{T%
}_{1-k,k,k}.
\end{equation}%
Taking into account (\ref{caritaredonda}), we notice that the perturbation $%
\mathcal{T}_{1-\frac{k}{2},\frac{k}{2},\frac{k}{2}}$ corresponds to an
operator that satisfies the unitarity bound $1-k<2j<-1$ only for $k>3$,
while the operator $\mathcal{T}_{1-k,k,k}$ does not satisfy that bound for
any value of $k$ grater than $2$. With operators (\ref{Considering}), we
define the following correlation functions 
\begin{equation*}
\left\langle \widetilde{\mathcal{T}}_{j_{1},m_{1},\overline{m}_{1}}(z_{1})...%
\widetilde{\mathcal{T}}_{j_{N},m_{N},\overline{m}_{N}}(z_{N})\right\rangle
_{S_{[\lambda ]}}=\frac{\Gamma (-s)}{b}\delta \left(
s+1+j_{1}+...j_{N}+M+(N-2-M)k/2\right) \times
\end{equation*}%
\begin{eqnarray}
&&\times \frac{1}{M!c_{k}^{M}}\delta \left( m_{1}+\overline{m}_{1}+...m_{N}+%
\overline{m}_{N}+k(M+2-N)\right) \delta _{m_{1}-\overline{m}_{1}+...+m_{N}-%
\overline{m}_{N}}\times  \notag \\
&&\times \prod_{r=1}^{M}\int d^{2}w_{r}\prod_{t=1}^{s}\int
d^{2}v_{t}\left\langle \widetilde{\mathcal{T}}_{j_{1},m_{1},\overline{m}%
_{1}}(z_{1})...\widetilde{\mathcal{T}}_{j_{N},m_{N},\overline{m}%
_{N}}(z_{N})\prod_{r=1}^{M}\mathcal{T}_{1-\frac{k}{2},\frac{k}{2},\frac{k}{2}%
}(w_{r})\prod_{t=1}^{s}\mathcal{T}_{1-k,k,k}(v_{t})\right\rangle
_{S_{[\lambda =0]}},  \label{yesto}
\end{eqnarray}%
where $b^{-2}=k-2$, and where we fixed $a_{k}=c_{k}^{-2}$. The value of $%
\lambda $ was also fixed to a specific value. The vertex operators $%
\widetilde{\mathcal{T}}_{j,m,\overline{m}}$ appearing in this expression are
related to those introduced in (\ref{caritaredonda}) through 
\begin{equation}
\widetilde{\mathcal{T}}_{j,m,\overline{m}}=\frac{c_{k}\Gamma (-m-j)}{\pi
^{2}\Gamma (1+j+\overline{m})}\mathcal{T}_{\widetilde{j},\widetilde{m},%
\widetilde{\overline{m}}},  \label{conjugate}
\end{equation}%
with 
\begin{equation}
\widetilde{j}=-j(k-1)-m(k-2)-k/2,\quad \widetilde{m}=jk+m(k-1)+k/2,
\label{j}
\end{equation}%
and analogously\footnote{%
Notice that it means that now anti-holomorphic contribution comes with a
different index $\widetilde{\overline{j}}$.} for $\widetilde{\overline{m}}.$
Again, notice that in (\ref{yesto}) we already fixed the value of $\lambda $
to a specific value $c_{k},$ which is a $k$-dependent numerical factor that
is ultimately related to the one appearing in (\ref{rrtt}). Realization (\ref%
{yesto}) is similar to (\ref{despues}) in Liouville field theory and defines
the correlators we want to consider here. The overall factor $\frac{\Gamma
(-s)}{bM!c_{k}^{M}}$ and the $\delta $-functions are understood once the
prescription for inserting the screenings when computing correlators is
specified. These factors come from the integration over the zero-modes of
the fields. Besides, the condition $\sum_{i=1}^{N}(m_{i}-\overline{m}_{i})=0$
also holds. The conditions imposed by these $\delta $-functions are
equivalent to demanding $\sum_{i=1}^{N}(\widetilde{m}_{i}-\widetilde{%
\overline{m}}_{i})=0$ and $\sum_{i=1}^{N}(\widetilde{m}_{i}+\widetilde{%
\overline{m}}_{i})+k(M+2s)=0,$ being $M+s$ the total amount of screenings to
be inserted. We discuss our prescription for the insertion of the screening
charges below.

\subsubsection{A Coulomb gas-like prescription}

In this realization, the interaction operators $\mathcal{T}_{1-\frac{k}{2},%
\frac{k}{2},\frac{k}{2}}$ and $\mathcal{T}_{1-k,k,k}$ act in (\ref{yesto})
as screening operators, analogously to the computation of Liouville
correlation functions. Because of the $\delta $-functions appearing in (\ref%
{yesto}), the amount of these screening operators to be inserted turns out
to be given by%
\begin{equation}
s=1-N-\sum_{i=1}^{N}j_{i}+\frac{k-2}{2}(M+2-N),\quad
M=N-2+\sum_{i=1}^{N}\omega _{i}.  \label{pupodos}
\end{equation}%
However, the statement is not complete unless one specifies how conditions (%
\ref{pupodos}) are to be satisfied. This is because in principle there is no
a unique way of choosing $s$ and $M$ in order to obey the first of the
charge symmetry conditions in (\ref{pupodos}). Thus, let us be precise about
the prescription to compute the r.h.s. of (\ref{yesto}): The prescription
adopted here is that $M$ represents a positive integer number of operators $%
\mathcal{T}_{1-\frac{k}{2},\frac{k}{2},\frac{k}{2}}$ to be inserted, and $M$
is actually fixed by the winding numbers $\omega _{i}$ of the $N$
interacting states. On the other hand, the amount $s$ of operators $\mathcal{%
T}_{1-k,k,k}$ is then appropriately chosen to make the r.h.s. of (\ref{yesto}%
) nonzero; and this is supposed to be the case even if (\ref{yesto}) has to
be analytically extended to non-integer values of $s$ (we already discussed
correlation functions with a non-integer amount of screening operators in
section 2). This is the set of correlation functions we will consider here;
and we emphasize that the equivalence between CFTs we will state in the
following subsection 4.2.3 has to be understood as holding only if the
prescription employed to compute the observables is the one we just gave in
this subsection. Now, once we precisely defined the correlators (\ref{yesto}%
), let us present the main assertion.

\subsubsection{Correspondence between correlation functions}

The statement is that the following identity between correlation functions
holds%
\begin{equation}
\left\langle \Phi _{j_{1},m_{1},\overline{m}_{1}}^{\omega
_{1}}(z_{1})...\Phi _{j_{N},m_{N},\overline{m}_{N}}^{\omega
_{N}}(z_{N})\right\rangle _{WZW}=\widehat{c}_{k}^{-2}\left\langle \widetilde{%
\mathcal{T}}_{j_{1},m_{1},\overline{m}_{1}}(z_{1})...\widetilde{\mathcal{T}}%
_{j_{N},m_{N},\overline{m}_{N}}(z_{N})\right\rangle _{S}  \label{duality}
\end{equation}%
where $\widehat{c}_{k}^{2}$ is a numerical factor (independent of $N$) that
will be specified below, and where the correlators in the r.h.s. are given
by (\ref{yesto}) computed with the prescription specified above. The average
in the r.h.s. of (\ref{duality}) is taken w.r.t. the action (\ref{S}) with $%
\lambda _{n}=0$ for $n\neq 1$ and $n\neq 2$. Recall%
\begin{equation}
\widetilde{\mathcal{T}}_{j,m,\overline{m}}=\frac{c_{k}\Gamma (-m-j)}{\pi
^{2}\Gamma (1+j+\overline{m})}\mathcal{T}_{\widetilde{j},\widetilde{m},%
\widetilde{\overline{m}}}.
\end{equation}

Relation (\ref{duality}) reads 
\begin{equation*}
\left\langle \Phi _{j_{1},m_{1},\overline{m}_{1}}^{\omega
_{1}}(z_{1})...\Phi _{j_{N},m_{N},\overline{m}_{N}}^{\omega
_{N}}(z_{N})\right\rangle _{WZW}=\frac{\Gamma (-s)}{\widehat{c}%
_{k}^{2}bM!c_{k}^{M-N}}\prod_{i=1}^{N}\frac{\Gamma (-m_{i}-j_{i})}{\Gamma
(1+j_{i}+\overline{m}_{i})}\delta _{m_{1}-\bar{m}_{1}+...+m_{N}-\bar{m}%
_{N}}\times
\end{equation*}%
\begin{equation*}
\times \delta \left( \sum_{i=1}^{N}(m_{i}+\overline{m}_{i})+(M+2-N)k\right)
\delta \left( s+1+\sum_{i=1}^{N}j_{i}+M+(N-2-M)k/2\right) \times
\end{equation*}%
\begin{equation}
\times \prod_{r=1}^{M}\int d^{2}w_{r}\prod_{t=1}^{s}\int
d^{2}v_{t}\left\langle \prod_{i=1}^{N}\mathcal{T}_{\widetilde{j}_{i},%
\widetilde{m}_{i},\widetilde{\overline{m}}_{i}}(z_{i})\prod_{r=1}^{M}%
\mathcal{T}_{1-\frac{k}{2},\frac{k}{2},\frac{k}{2}}(w_{r})\prod_{t=1}^{s}%
\mathcal{T}_{1-k,k,k}(v_{t})\right\rangle _{S_{[\lambda =0]}}.
\label{duality2}
\end{equation}%
This is the main result here. Equation (\ref{duality}) gives a realization
of any $N$-point function of the $SL(2,\mathbb{R})_{k}/U(1)$ WZW correlators
in terms of the analogous observables in the theory (\ref{S}) if the
perturbation is taken to be $\lambda _{n}=\mu \delta _{n-2}+\lambda \delta
_{n-1}$ (and choosing $\mu $ and $\lambda $ appropriately). The perturbation
involved in this realization corresponds to the operators $\mathcal{T}_{1-%
\frac{k}{2},\frac{k}{2},\frac{k}{2}}$ and $\mathcal{T}_{1-k,k,k},$ having
momentum modes $n=1$ and $n=2,$ respectively. This is different from the
standard FZZ duality, which corresponds to $\lambda _{n}=\widetilde{\lambda }%
\delta _{n-1}+\widetilde{\lambda }\delta _{n+1}$, instead. The perturbation
of the linear dilaton theory (\ref{S0}) with operators of different winding
numbers was also considered in Ref. \cite{MMP}, where it was suggested that
the multiply-wound tachyon operators are linked to the called higher-spin
black holes. It would be very interesting to understand the relation with
the realization of \cite{MMP} better and confirm such picture.

\subsubsection{Conjugate representations and spectral flow}

An interesting feature of the statement made above is that the r.h.s. of (%
\ref{duality}) involves \textquotedblleft conjugate operators" instead of
the ones introduced in (\ref{caritaredonda}). Ones are related to each
others by (\ref{j}), which represents a symmetry of the formula for the
conformal dimension (\ref{h}) (and not only for it, actually). Notice that,
in particular, we have 
\begin{equation}
\widetilde{\mathcal{T}}_{1-\frac{k}{2},\frac{k}{2},\frac{k}{2}}\sim \mathcal{%
T}_{1-k,k,k},
\end{equation}%
and operators $\widetilde{\mathcal{T}}_{j,m,\overline{m}}$ and $\mathcal{T}%
_{j,m,\overline{m}}$ have exactly the same conformal dimension. Moreover,
the automorphism can be extended in order to be valid for the theory
formulated on the product $SL(2,\mathbb{R})_{k}/U(1)\times \mathbb{R} $ by
including the new winding number 
\begin{equation}
\widetilde{\omega }=-\omega -1-2(j+m).  \label{ww}
\end{equation}%
In such case, the operators $\widetilde{\mathcal{T}}_{j,m,\overline{m}}$ and 
$\mathcal{T}_{j,m,\overline{m}}$, when both are extended by including the
time-like factor $e^{i\sqrt{\frac{2}{k}}(m+\frac{k}{2}\omega )T}$, satisfy%
\begin{equation*}
-\frac{j(j+1)}{k-2}-m\omega -\frac{k}{4}\omega ^{2}=-\frac{\widetilde{j}(%
\widetilde{j}+1)}{k-2}-\widetilde{m}\widetilde{\omega }-\frac{k}{4}%
\widetilde{\omega }^{2}
\end{equation*}%
and also have the same momentum under the $J^{3}$ current of the WZW model,
namely%
\begin{equation*}
m+\frac{k}{2}\omega =\widetilde{m}+\frac{k}{2}\widetilde{\omega }.
\end{equation*}
To understand the relation between $\widetilde{\mathcal{T}}_{j,m,\overline{m}%
}$ and $\mathcal{T}_{j,m,\overline{m}}$ in the algebraic framework, let us
comment on the $SL(2,\mathbb{R})_{k}$ representations again: As we said, the
principal continuous series $\mathcal{C}_{\lambda }^{\alpha ,\omega }$
correspond to $j=-\frac{1}{2}+i\lambda $ with $\lambda \in \mathbb{R}$ and
thus, through (\ref{j}), this results in the new values $j=-\frac{1}{2}+i%
\tilde{\lambda}-m(k-2)$ with $\tilde{\lambda}\in \mathbb{R}$, which only
belong to the continuous series if $m=0$. Besides, if we perform the change (%
\ref{j}) for generic $\tilde{\lambda}$, then $m$ turns out to be a non-real
number after of that. Then, the relation between $j,m$ and $\widetilde{j},%
\widetilde{m}$ can not be thought of as a simple identification between
states of different continuous representations but it does correspond to
different free field realizations (at least in what respects to the
continuous series $\mathcal{C}_{\lambda }^{\alpha ,\omega }$). On the other
hand, concerning the discrete representations, it is worth mentioning that
the quantity $j+m$ remains invariant under the involution (\ref{j}); though
it is not the case for the difference $j-m$ that, instead, remains invariant
under a $\mathbb{Z}_{2}$ reflected version of (\ref{j}). Then, unlike the
states of continuous representations, the transformation defined by (\ref{j}%
) and (\ref{ww}) is closed among certain subset of states of discrete
representations. This is because such transformation maps states of the
discrete series with $2(j+m)\in \mathbb{Z}$ among themselves. In particular,
the case $m+j=0$ corresponds to the well known identification between
discrete series $\mathcal{D}_{j}^{\pm ,\omega =0}$ and $\mathcal{D}%
_{-k/2-j}^{\mp ,\omega =\pm 1}$ since in that case (\ref{j}) and (\ref{ww})
reduce to $j\rightarrow -k/2-j,\ m\rightarrow k/2-m=k/2+j,\ \omega
\rightarrow -1-\omega $ (i.e. it includes\footnote{%
More precisely, the identification between Kac-Moody primary highest-weight
(lowets-weight) states that is induced by the sector $\omega =1$ of the
spectral flow coincides with a particular case of the identification given
by the symmetry (\ref{j}), (\ref{ww}).} such spectral flow transformation as
a particular case). Also notice that the condition $m-\bar{m} \in \mathbb{Z}$
is not preserved for generic values of $k$. The fixed points of (\ref{j})
describe a line in the space of representations, parameterized by $%
j+1/2=-m(k-2)/k$; in particular, a fixed point for generic $k$ corresponds
to $j=-1/2$ and $m=0 $, for which (\ref{ww}) reduces to $\omega \rightarrow
-\omega $. Also, in the tensionless limit $k\rightarrow 2$ transformation (%
\ref{j}) agrees with the Weyl reflection $j\rightarrow -1-j$. The relation
between quantum numbers manifested by (\ref{j}) permits to visualize the
relation between the vertex considered in our construction and those of
reference \cite{HHS}, and we emphasize that these correspond to two
different (alterative) representations of the vertex operators. The relation
between both is a kind of \textquotedblleft twisting\textquotedblright\ and
is presumably related to the representations studied in \cite{GL} for the
WZW theory. Certainly, the conjugate representations of the $SL(2,\mathbb{R}%
)_{k}$ vertex algebra do resemble the twisting (\ref{conjugate}), and it
seems to be a nice connection between correlators (\ref{yesto}) and the free
field representation studied in \cite{GN2,GN3,GL,HHS}. Conjugate
representations transform in a particular way under the Kac-Moody affine $%
\widehat{sl}(2)_{k} $ algebra, and are analogous to those introduced by
Dotsenko for the case of $\widehat{su}(2)_{k}$ in Refs. \cite{dot1,dot2}.
For the non-compact WZW these were first introduced in Ref. \cite{FZZ} to
describe winding violating amplitudes. Here, through Eq. (\ref{conjugate}),
these appear again (although we are referring to them as \textquotedblleft
twisted") within a similar context.

\subsubsection{Remark on the $\hat{sl}(2)_{k}$ affine symmetry}

To understand these twisted sectors better, let us make some remarks on the $%
sl(2)_{k}$ symmetry of the action (\ref{S}) when it is perturbed as we did
so. We are claiming (and we will prove in the following subsection) that
correlators (\ref{yesto}) do transform appropriately under the $sl(2)_{k}$
symmetry in order to describe the WZW correlators. However, even though one
can prove this a posteriori, the question arises as to why does it happen if
the operators $\mathcal{T}_{1-k,k,k}$ do not seem to commute with the $%
\widehat{sl}(2)_{k}$ currents though. To be precise, even though one
eventually proves that the free field representation employed here
transforms properly by construction (\textit{e.g.} it reproduces solutions
of the KZ equation), it is also true that this is not obvious because the
screening operators do not seem to commute with the free field
representation of the $sl(2)_{k}$ current algebra (\ref{currents}) as one
could naively expect. The explanation of this puzzling feature is that the
vertex operators $\widetilde{\mathcal{T}}_{j,m,\overline{m}}$ do not satisfy
the usual OPE with the $sl(2)_{k}$ currents either, and thus this restores
the symmetry. To see this explicitly, one has to consider the generators of
the affine algebra (\ref{currents}) and verify that those currents do not
have regular OPE with the operators $\mathcal{T}_{1-k,k,k}$. The remarkable
point is that this is precisely what makes the $SL(2,\mathbb{R})_{k}$ to be
restored: While these currents do not present regular OPE with the operator $%
\mathcal{T}_{1-k,k,k}$, these do not satisfy the usual OPE with the twisted
vertex operators $\widetilde{\mathcal{T}}_{j,m,\overline{m}}$ either; and
both facts seem to combine in such a way that render the set of observables (%
\ref{yesto}) $SL(2,\mathbb{R})_{k}$ invariant\footnote{%
G.G. thanks Yu Nakayama for addressing his attention to this remarkable
point.}. This depends on the presence of the normalization factor $\frac{%
\Gamma (-j-m)}{\Gamma (j+\overline{m}+1)}$ in (\ref{conjugate}), since its
presence is not innocuous for the transformation properties under the
generators $J_{n}^{\pm }$. This feature makes out of the correspondence (\ref%
{duality}) a non-trivial assertion, indeed.

\subsection{Proving the correspondence}

Here we will show that the formula (\ref{duality}) immediately follows from
the relation (\ref{rrtt}) between WZW and Liouville correlators. In order to
be concise, here we address the proof in two steps: First, we rewrite the
correlators (\ref{yesto}) and the operators involved there in a convenient
way\footnote{%
In Ref. \cite{Fateev} similar techniques were used to prove a different
(though related) correspondence: the one between Liouville and sine-Liouvile
correlation functions.}. The second step will be using the formula (\ref%
{rrtt}) to make contact with the WZW correlators.

\subsubsection{Step 1: Rewriting the correlators}

As we said, the proof of formula (\ref{duality}) directly follows from the
Stoyanovsky-Ribault-Teschner map (\ref{rrtt}) we discussed in section 2. In
order to make the proof simpler, let us begin by redefining fields as follows

\begin{equation}
\varphi (z)=(1-k)\widehat{\varphi }(z)+i\sqrt{k(k-2)}\widehat{X}(z),\quad
X(z)=i\sqrt{k(k-2)}\widehat{\varphi }(z)+(k-1)\widehat{X}(z).  \label{newc}
\end{equation}%
That is 
\begin{equation}
-\sqrt{\frac{k-2}{2}}\widehat{\varphi }+i\sqrt{\frac{k}{2}}\widehat{X}=-%
\sqrt{\frac{k-2}{2}}\varphi +i\sqrt{\frac{k}{2}}X  \label{Argento2}
\end{equation}%
and%
\begin{equation}
-\frac{1}{\sqrt{k-2}}\varphi =\frac{k-1}{\sqrt{k-2}}\widehat{\varphi }-i%
\sqrt{k}\widehat{X}.  \label{Argento1}
\end{equation}%
Also notice that it implies 
\begin{equation}
\partial \varphi \overline{\partial }\varphi +\partial X\overline{\partial }%
X=\partial \widehat{\varphi }\overline{\partial }\widehat{\varphi }+\partial 
\widehat{X}\overline{\partial }\widehat{X};  \label{Argento3}
\end{equation}%
so that the free field correlators are $\left\langle \widehat{\varphi }%
(z_{1})\widehat{\varphi }(z_{2})\right\rangle =\left\langle \widehat{X}%
(z_{1})\widehat{X}(z_{2})\right\rangle =-2\log |z_{1}-z_{2}|$. One can
wonder whether the field redefinition (\ref{newc}) is well defined or not
since it is a complex transformation and then both $\widehat{\varphi }$ and $%
\widehat{X}$ would acquire a non-real part. However, the correct way of
thinking this transformation is first considering a Wick rotation of the $X$
direction and then, after the transformation, Wick rotate $\widehat{X}$
back. It turns out to be a perfectly defined transformation for the Wick
rotated fields $iX$ and $i\widehat{X}$, which can be seen as real time-like
bosons. Transformation (\ref{newc}) is a $U(1,1)$ transformation, with
determinant $-1$. In fact, one can also turn it into a $SU(2)$-rotation by
supplementing (\ref{newc}) with a reflection $X\rightarrow -X$ (that is also
a symmetry of the theory). In that case, it is clear that (\ref{Argento1})
and (\ref{Argento3}) remain invariant, while (\ref{Argento2}) changes its
sign in the second term of the r.h.s.. So, in principle, it would be
possible to consider a $U(1,1)\times \overline{SU}(2)$ chiral transformation
(for the holomorphic part and the anti-holomorphic part, respectively) in
order to transform dependences on $X$ into dependences on $\widetilde{X}$.

In terms of these new fields $\widehat{\varphi }$ and $\widehat{X}$ one
finds that the linear dilaton theory defined by the action $S_{0}-\frac{1}{%
4\pi }\int d^{2}z\ \partial T\bar{\partial}T$ takes the form%
\begin{equation}
S=\frac{1}{4\pi }\int d^{2}z\left( -\partial T\bar{\partial}T+\partial 
\widehat{X}\bar{\partial}\widehat{X}+\partial \widehat{\varphi }\bar{\partial%
}\widehat{\varphi }+\frac{1}{2\sqrt{2}}QR\widehat{\varphi }-\frac{i}{2}\sqrt{%
\frac{k}{2}}R\widehat{X}\right)  \label{SA}
\end{equation}%
with $Q=b+b^{-1}$, and $b^{-2}=k-2$, so that $Q=\frac{k-1}{\sqrt{(k-2)}}%
=b+b^{-1}$ (cf. Eq. (\ref{T}) in section 2). That is, the background charge
operator $e^{-\sqrt{\frac{2}{k-2}}\varphi }$ transform through (\ref{j})
into a new background charge operator $e^{\sqrt{2}Q\varphi -i\sqrt{2k}X}$,
where $\tilde{j}=-1$, $\tilde{m}=0$ while $j=k-1$, $m=-k$. Consequently, the
stress-tensor reads \cite{0511252} 
\begin{equation}
T(z)=\frac{1}{2}(\partial T)^{2}-\frac{1}{2}(\partial \widehat{X})^{2}-i%
\sqrt{\frac{k}{2}}\partial ^{2}\widehat{X}-\frac{1}{2}(\partial \widehat{%
\varphi })^{2}+\frac{k-1}{\sqrt{2(k-2)}}\partial ^{2}\widehat{\varphi },
\end{equation}%
and the dilaton now acquires a linear dependence on both directions $%
\widehat{X}$ and $\widehat{\varphi }$. This kind of CFT, representing $c<1$
matter coupled to perturbed 2D gravity, was recently discussed in Refs. \cite%
{Petkova,Petkova2,ZamolodchikovYang,BelavinZamolodchikov,Z}. It is possible
to verify that this stress-tensor leads to the appropriated central charge $%
c=3+\frac{6}{k-2},$ as it is of course expected. On the other hand, in terms
of the new fields the interaction (perturbation) term $\lambda \mathcal{T}%
_{1-\frac{k}{2},\frac{k}{2},\frac{k}{2}}(w_{r})+\mu \mathcal{T}%
_{1-k,k,k}(v_{t})$ takes the form%
\begin{equation}
\mathcal{O}_{\lambda _{1},\lambda _{2}}=c_{k}^{-1}e^{-\sqrt{\frac{k-2}{2}}%
\widehat{\varphi }+i\sqrt{\frac{k}{2}}\widehat{X}}+e^{\sqrt{\frac{2}{k-2}}%
\widehat{\varphi }},  \label{SB}
\end{equation}%
where we already fixed the scale $\mu $ to a specific value by shifting the
zero-mode of the Liouville field $\widehat{\varphi }$, and we also specified
the numerical factor $c_{k}$ as being the ratio between the couplings $\mu $
and $\lambda ^{2}$ in (\ref{Uf}). Notice that in these coordinates the
second term in the perturbation $\mathcal{O}_{\lambda _{2},\lambda _{1}}$
turns to be diagonalized (no dependence on $\widehat{X}$ arise there) and
agrees with the Liouville cosmological constant $\mu e^{\sqrt{2}b\widehat{%
\varphi }}$. On the other hand, the first term in (\ref{SB}) still has the
form of one of the two exponentials that form the cosine interaction (\ref%
{cos2})\ in sine-Liouville theory; this is due to (\ref{Argento2}). On the
other hand, the vertex operators in terms of $\widehat{X}$ and $\widehat{%
\varphi }$ take the form\footnote{%
Again, we are not explicitly writing the antiholomorphic contribution $e^{i%
\sqrt{\frac{2}{k}}\overline{m}\widehat{X}}$ for short. It has to be
understood in what follows.}%
\begin{equation}
\widetilde{\mathcal{T}}_{j,m,\overline{m}}=\frac{c_{k}\Gamma (-m-j)}{\pi
^{2}\Gamma (1+\overline{m}+j)}V_{\alpha }\times e^{i\sqrt{\frac{2}{k}}(m-%
\frac{k}{2})\widehat{X}+i\sqrt{\frac{2}{k}}(m+\frac{k}{2}\omega )T},
\label{oh}
\end{equation}%
with the Liouville field $V_{\alpha }=e^{\sqrt{2}\alpha \widehat{\varphi }}$%
, with $\alpha =bj+b+b^{-2}/2=b(j+k/2).$ Expanding the correlators we get 
\begin{equation*}
\frac{1}{\hat{c}_{k}^{2}}\left\langle \widetilde{\mathcal{T}}_{j_{1},m_{1},%
\overline{m}_{1}}(z_{1})...\widetilde{\mathcal{T}}_{j_{N},m_{N},\overline{m}%
_{N}}(z_{N})\right\rangle _{S_{[\lambda ,\mu ]}}=\frac{\Gamma (-s)}{%
bM!c_{k}^{M}\widehat{c}_{k}^{2}}\delta _{m_{1}-\overline{m}_{1}+..m_{N}-%
\overline{m}_{N}}\times
\end{equation*}%
\begin{eqnarray*}
&&\times \delta \left( s+1+j_{1}+...j_{N}+M+(N-2-M)k/2\right) \delta \left(
m_{1}+\overline{m}_{1}+..m_{N}+\overline{m}_{N}+(N-2-M)k\right) \times \\
&&\times \prod_{r=1}^{M}\int d^{2}w_{r}\prod_{t=1}^{s}\int
d^{2}v_{t}\left\langle \prod_{i=1}^{N}\widetilde{\mathcal{T}}_{j_{i},m_{i},%
\overline{m}_{i}}(z_{i})\times \prod_{r=1}^{M}\mathcal{T}_{1-\frac{k}{2},%
\frac{k}{2},\frac{k}{2}}(w_{r})\prod_{t=1}^{s}\mathcal{T}_{1-k,k,k}(v_{t})%
\right\rangle _{S_{[\lambda =0,\mu =0]}}
\end{eqnarray*}%
and this can be written as 
\begin{eqnarray*}
&=&\frac{1}{k\hat{c}_{k}^{2}}\delta \left( \omega _{1}+...\omega
_{N}+N-2-M\right) \prod_{a=1}^{N}\frac{c_{k}\ \Gamma (m_{a}-j_{a})}{\pi
^{2}\Gamma (1+j_{a}-\bar{m}_{a})}\left\langle \prod_{t=1}^{N}e^{i\sqrt{\frac{%
2}{k}}(m_{t}+\frac{k}{2}\omega _{t})T(z_{t})}\right\rangle _{S_{[\lambda
=0]}}\times \\
&&\times \frac{1}{M!c_{k}^{M}}\delta _{m_{1}-\overline{m}_{1}+...m_{N}-%
\overline{m}_{N}}\prod_{r=1}^{M}\int d^{2}w_{r}\ \left\langle
\prod_{t=1}^{N}e^{i\sqrt{\frac{2}{k}}(m_{t}-\frac{k}{2})\widehat{X}%
(z_{t})}\prod_{r=1}^{M}e^{i\sqrt{\frac{k}{2}}\widehat{X}(w_{r})}\right%
\rangle _{S_{[\lambda =0]}}\times
\end{eqnarray*}%
\begin{equation}
\times \frac{\Gamma (-s)}{b}\delta \left( s-1-\frac{2+M}{2b^{2}}+\frac{%
\alpha _{1}+...\alpha _{N}}{b}\right) \prod_{t=1}^{s}\int d^{2}v_{t}\
\left\langle \prod_{t=1}^{N}V_{\alpha _{t}}(z_{t})\prod_{r=1}^{M}V_{-\frac{1%
}{2b}}(w_{r})\prod_{t=1}^{s}V_{b}(v_{t})\right\rangle _{S_{L[\mu =0]}},
\label{dalegas}
\end{equation}%
where $S_{[\lambda =0]}$ here refers to the unperturbed action%
\begin{equation}
S_{[\lambda =0]}=\frac{1}{4\pi }\int d^{2}z\left( -\partial T\overline{%
\partial }T+\partial \widehat{X}\overline{\partial }\widehat{X}-\frac{i}{2}%
\sqrt{\frac{k}{2}}R\widehat{X}\right) .  \label{uhyeste}
\end{equation}%
The third line in (\ref{dalegas}) turns out to be a $N+M$-point correlation
function in Liouville field theory (see Eq. (\ref{despues}) in section 2),
defined by the Liouville action%
\begin{equation*}
S_{L}[\mu ]=\frac{1}{4\pi }\int d^{2}z\left( \partial \widehat{\varphi }%
\overline{\partial }\widehat{\varphi }+\frac{1}{2\sqrt{2}}QR\widehat{\varphi 
}+2\pi \mu e^{\sqrt{2}b\widehat{\varphi }}\right) ,
\end{equation*}%
with $Q=b+b^{-1}$. Recall that the parameter $b$ of the Liouville theory is
related to the Kac-Moody level $k$ through $b^{-2}=k-2$, while the quantum
numbers $\alpha _{i}$ are defined in terms of $j_{i}$ by $\alpha
_{i}=bj_{i}+b+b^{-2}/2,$ for$\ i=1,2,...N.$ After the Wick contraction, we
find 
\begin{equation*}
\frac{\Gamma (-s)}{bM!c_{k}^{M}\widehat{c}_{k}^{2}}\prod_{r=1}^{M}\int
d^{2}w_{r}\prod_{t=1}^{s}\int d^{2}v_{t}\left\langle \widetilde{\mathcal{T}}%
_{j_{1},m_{1},\overline{m}_{1}}(z_{1})...\widetilde{\mathcal{T}}%
_{j_{N},m_{N},\overline{m}_{N}}(z_{N})\prod_{r=1}^{M}\mathcal{T}_{1-\frac{k}{%
2},\frac{k}{2},\frac{k}{2}}(w_{r})\prod_{t=1}^{s}\mathcal{T}%
_{1-k,k,k}(v_{t})\right\rangle _{S[\lambda =0]}=
\end{equation*}%
\begin{equation}
=N_{k}(j_{1},...j_{N};m_{1},...m_{N})\prod_{r=1}^{M}\int d^{2}w_{r}\
F_{k}(z_{1},...z_{N};w_{1},...w_{M})\langle \prod_{t=1}^{N}V_{\alpha
_{t}}(z_{t})\prod_{r=1}^{M}V_{-\frac{1}{2b}}(w_{r})\rangle _{S_{L}[\mu ]}\ ,
\label{rt}
\end{equation}%
where $\mu =b^{2}/\pi ^{2}$ and where, after fixing the value $\hat{c}%
_{k}^{2}=2c_{k}^{2}/b\pi ^{3}$, the normalization factor is 
\begin{equation}
N_{k}(j_{1},...j_{N};m_{1},...m_{N})=\frac{2\pi ^{3-2N}b}{M!\ c_{k}^{M+2-N}}%
\prod_{i=1}^{N}\frac{\Gamma (-m_{i}-j_{i})}{\Gamma (1+j_{i}+\bar{m}_{i})}
\end{equation}%
and the function $F_{k}(z_{1},...z_{N};w_{1},...w_{M})$ is given by 
\begin{eqnarray}
F_{k}(z_{1},...z_{N};w_{1},...w_{M}) &=&\frac{\prod_{1\leq
r<l}^{N}|z_{r}-z_{l}|^{k-2(m_{r}+m_{l}+\omega _{r}\omega _{l}k/2+\omega
_{l}m_{r}+\omega _{r}m_{l})}}{\prod_{1<r<l}^{M}|w_{r}-w_{l}|^{-k}%
\prod_{t=1}^{N}\prod_{r=1}^{M}|w_{r}-z_{t}|^{k-2m_{t}}}\times  \notag \\
&&\times \frac{\prod_{1\leq r<l}^{N}(\bar{z}_{r}-\bar{z}_{l})^{m_{r}+m_{l}-%
\bar{m}_{r}-\bar{m}_{l}+\omega _{l}(m_{r}-\bar{m}_{r})+\omega _{r}(m_{l}-%
\bar{m}_{l})}}{\prod_{1<r<l}^{M}(\bar{w}_{r}-\bar{z}_{t})^{m_{t}-\bar{m}_{t}}%
}.  \label{F}
\end{eqnarray}%
Remarkably, this has reproduced the r.h.s. of formula (\ref{rrtt}); cf. Eq. (%
\ref{FF}). Notice that the exponents of the differences $|z_{r}-z_{l}|$ in (%
\ref{F}) do depend on whether the theory is being formulated on the coset $%
SL(2,\mathbb{R})_{k}/U(1)$ or on its product with the time $T$. The vertex
operators (\ref{oh}) are the only fields that carry the $T$-dependences, so
that the rest of the OPEs are not affected.

According to (\ref{pupodos}), the amount of perturbations involved in (\ref%
{rt}) is constrained by the following conditions 
\begin{equation}
\sum_{i=1}^{N}m_{i}=\sum_{i=1}^{N}\bar{m}_{i}=\frac{k}{2}(N-M-2),\quad
s=-b^{-1}\sum_{i=1}^{N}\alpha _{i}+b^{-2}\frac{M}{2}+1+b^{-2}\ ,  \label{ese}
\end{equation}%
where the number $s$ corresponds to the amount of screening operators $%
V_{b}=\mu e^{\sqrt{2}b\widehat{\varphi }}$ to be included in Liouville
correlators. The whole amount of vertex operators involved in the r.h.s. of (%
\ref{rt}) is then $N+M+s,$ and is related to the winding numbers of the
strings through 
\begin{equation}
\sum_{i=1}^{N}\omega _{i}=M+2-N\geq -|N-2|.  \label{selection}
\end{equation}%
Notice that the value of $\sum_{i=1}^{N}\omega _{i}$ can not be lower than $%
2-N$ if $M$ represents a positive integer number. Allowing negative values
of winding numbers requires the insertion of screening operators with $n=-1$
in addition to those of $n=+1$. $M$ runs between $0 $ and $N-2,$ which
implies that, according to the prescription given in subsection 4.2.2, the
absolute value of the violation of winding number conservation could not
exceed $N-2$. This is an interesting feature, and it is not trivial at all
to fully understand this bound. We can say here that it is closely related
to the $\widehat{sl}(2)_{k}$ symmetry of the theory, and we refer to the
appendix D of Ref. \cite{MO3} for a nice explanation. It is worth mentioning
that the selection rule for winding number violation (\ref{selection}) was
already part of the original FZZ conjecture \cite{FZZ}. A short note about
this rule can be also found in Ref. \cite{Fateev3}.

The formula (\ref{rt})-(\ref{F}), with the conditions (\ref{ese}), is the
main ingredient for proving (\ref{duality}). It only remains to argue that
the r.h.s. of (\ref{rt}) actually represents a WZW correlator; and,
actually, it can be already observed since it directly follows from the
formula (\ref{rrtt}). Indeed, the r.h.s. of (\ref{rt}) agrees with the
l.h.s. of (\ref{rrtt}) and this would complete the proof of (\ref{duality}).
Let us conclude the job by further commenting on it.

\subsubsection{Step 2: Realizing the Stoyanovsky-Ribault-Teschner map}

As we just mentioned, the last step in proving (\ref{duality}) is showing
that the r.h.s. of Eq. (\ref{rt}) precisely describes a WZW $N$-point
function, and, actually, this immediately follows from the main result of
Ref. \cite{R} (see formula (3.29) there, which we wrote in the Eq. (\ref%
{rrtt}) in section 2). Hence, we have managed to rewrote our result (\ref%
{duality}) in such a way that its proof turns out to be a direct consequence
of the observation made by S. Ribault in his paper \cite{R}, where he showed
that the l.h.s. of Eq. (\ref{rt}) is precisely equal to a correlation
function in the $SL(2,\mathbb{R})_{k}$ WZW model. Our achievement was to
prove that the auxiliary overall function $%
F_{k}(z_{1},...z_{N};w_{1},...w_{M})$ standing in the Ribault-Teschner
formula can be also thought of as coming from the correlation functions of
the linear dilaton CFT realized by the field $\widehat{X}$; namely%
\begin{equation}
F_{k}(z_{1},...z_{N};w_{1},...w_{M})=\left\langle \prod_{t=1}^{N}e^{i\sqrt{%
\frac{2}{k}}\left( (m_{t}-\frac{k}{2})\widehat{X}(z_{t})+(m_{t}+\frac{k}{2}%
\omega _{t})T(z_{t})\right) }\prod_{r=1}^{M}e^{i\sqrt{\frac{k}{2}}\widehat{X}%
(w_{r})}\right\rangle _{S_{[\lambda =0]}}.  \label{torbellino}
\end{equation}%
That is, we showed how the Ribault-Teschner formula can be seen as an
identity between correlators of two different two-dimensional $\sigma $%
-models with three-dimensional target space each. While one of these is the $%
SL(2,\mathbb{R})_{k}$ WZW, the other is of the form%
\begin{equation}
Liouville\times {\mathcal{M}}_{k}\times time  \label{cardinate}
\end{equation}%
of which Liouville theory is just a part. The $\mathbb{R}$ factor
corresponds to the time-like direction, parameterized by $T$. On the other
hand, the ${\mathcal{M}}_{k}$ factor is a $U(1)$\ direction, parameterized
by the field $\widehat{X}$, and describes a linear dilaton theory with
central charge $c=1-6k<1$. In fact, notice that the contribution of $%
\widehat{X}$ to the central charge is actually negative because $k>2$. The
field $\widehat{X}$ interacts with the Liouville field $\widehat{\varphi }$
through the tachyon-like potential, so that the first product in (\ref%
{cardinate}), unlike the second, is not a \textit{direct} product. The time
direction, instead, does not interact with the other fields, and it only
contributes to the total central charge and the conformal dimension of the
vertex operators.

The fact that a construction like (\ref{cardinate}) is possible is not a
minor detail: Realizing that the Ribault-Teschner formula (\ref{rrtt})
admits to be interpreted as the equivalence between these two CFTs demanded
not only the existence of a realization like (\ref{torbellino}), but also
demanded the contribution of the central charge coming from the $U(1)\times 
\mathbb{R}$ part to agree with the difference between the Liouville central
charge $c_{L}=1+6Q^{2}$ and the $SL(2,\mathbb{R})$ WZW central charge $%
c_{SL(2)}=3+6\widehat{Q}^{2}=3k/(k-2)$, being reminded of $b^{-2}=k-2$.
Moreover, such value for the central charge of the CFT defined by fields $%
\widehat{X}$ and $T$ had to be consistent with the conformal dimension of
the fields in (\ref{torbellino}), leading to reproduce the formula for the
conformal dimension of the WZW vertex operators. Besides, another feature
that had to be explained was the presence of the $M$ additional fields $%
V_{-1/2b}$ arising in the r.h.s. of (\ref{rt}). Their presence is now
understood as follows: Since we know that the (the $M$-multiple integral of
the) product between the function $F_{k}(z_{1},...z_{N};w_{1},...w_{M})$ and
the $N+M$-point Liouville correlation function does satisfy the KZ\ equation
and so represents a correlation function in the WZW theory, it is then
expected that the Liouville degenerate fields $V_{-1/2b}$ arising there
admit to be expressed as a ($1,1$)-operator in a \textquotedblleft bigger"
theory with the form $Liouville\times CFT$ (i.e. the screening charge $%
V_{-1/2b}\times e^{i\sqrt{k/2}\widehat{X}}$ standing as the first term of
the r.h.s. of (\ref{SB})). That is, even though $V_{-1/2b}$ has dimension $%
h=-\frac{1}{2}-\frac{3}{4b^{2}}\neq 1$ with respect to the Liouville
stress-tensor, it does correspond to a ($1,1$)-operator\footnote{%
Even though the operator $V_{-1/2b}\times e^{i\sqrt{k/2}\widehat{X}}$ has
dimension $1$, it is not strictly correct to refer to it as a
\textquotedblleft screening" operator due to the remark on the $\widehat{sl}%
(2)_{k}$ transformation properties made in section 4.2.5.} $V_{-1/2b}\otimes
V_{CFT}$ with respect to the stress-tensor of the bigger model $%
Liouville\times CFT$. Of course, the theory also admits as a screening
operator the one that was already the screening for the \textquotedblleft
Liouville part of the theory", namely $V_{b}\otimes I$; so (\ref{SB}) can be
written as the sum of both, $\mathcal{O}_{\lambda _{1},\lambda
_{2}}=c_{k}^{-1}V_{-1/2b}e^{i\sqrt{k/2}\widehat{X}}+V_{b}$.

Notice that all the requirements mentioned above are actually obeyed by the
theory defined by the action (\ref{uhyeste}) perturbed by the operator (\ref%
{SB}). Hence, we have given a free field representation of the
Ribault-Teschner formula (\ref{rrtt}). Related to this, in Ref. \cite{R} it
was commented that a parafermionic realization of (\ref{rrtt}) is also
known, and the unpublished work by V. Fateev was referred. The parafermion
representation leads (see Eq. (3.31) in \cite{R}) to a formula similar to (%
\ref{rrtt}) provided the replacement of the factor $\prod_{1\leq
r<l}^{N}(z_{r}-z_{l})^{\frac{k}{2}-(m_{r}+m_{l}+\omega _{r}\omega _{l}\frac{k%
}{2}+\omega _{l}m_{r}+\omega _{r}m_{l})}(\overline{z}_{r}-\overline{z}_{l})^{%
\frac{k}{2}-(\overline{m}_{r}+\overline{m}_{l}+\omega _{r}\omega _{l}\frac{k%
}{2}+\omega _{l}\overline{m}_{r}+\omega _{r}\overline{m}_{l})}$ in (\ref{FF}%
) by a factor $\prod_{1\leq r<l}^{N}(z_{r}-z_{l})^{\frac{k}{2}+\frac{2}{k}%
m_{r}m_{l}-m_{r}-m_{l}}(\overline{z}_{r}-\overline{z}_{l})^{\frac{k}{2}+%
\frac{2}{k}\overline{m}_{r}\overline{m}_{l}-\overline{m}_{r}-\overline{m}%
_{l}},$ and notice that this is exactly what we find in our language (\ref%
{torbellino}) if we exclude the $T$ dependence in the vertex operators. This
realizes a correspondence like (\ref{rrtt}) but for the case of the coset $%
SL(2,\mathbb{R})_{k}/U(1)$. See the \textquotedblleft notes" at the end of
Ref. \cite{0511252} where the similarities with Fateev's work were already
mentioned. Besides, a realization of the Ribault-Teschner formula in terms
of Liouville times a $c<1$ matter CFT was independently presented by S.
Nakamura and V. Niarchos in Ref. \cite{Nuevo}. We would like to explore the
similarities between our realization and the one in that paper; we just
realized that the realization in \cite{Nuevo} does closely parallels ours.

Summarizing: because of Ribault-Teschner formula, it turns out that the
correlation function in the r.h.s. of (\ref{rt}) does correspond to the
string amplitude in the black hole ($\times time$) background, where the
winding number conservation is being violated in an amount $|N-2-M|$.\
Consequently, this implies that the l.h.s. of (\ref{rt}) do correspond to
WZW correlators as well, and this completes the proof of (\ref{duality}).
However, it has to be emphasized that the correspondence between BPZ and KZ
equations was proven for the Lorentzian theory, namely holding for
continuous representations. Thus, considering its validity beyond such
regime assumes a sort of analytic continuation. The convergence of integrals
in (\ref{rt}) is the subtle point here.

\subsection{A consistency check of the correspondence}

We have proven formula (\ref{duality}); this was first done in Ref. \cite%
{0511252}, but the order of the presentation was rather different there.
Formula (\ref{duality}) turns out to be a useful tool for computing
correlators in the WZW theory. A concise example was given in Ref. \cite%
{gaston}, where the free field representation in terms of the $%
Liouville\times U(1)\times time$ conformal field theory (\ref{SA})-(\ref{SB}%
) was employed to compute WZW three-point functions for the particular case
where the total winding number is violated in one unit. This quantity turns
out to be proportional to the Liouville correlator%
\begin{equation*}
\left\langle \Phi _{j_{1},m_{1},\overline{m}_{1}}^{\omega _{1}}(0)\Phi
_{j_{2},m_{2},\overline{m}_{2}}^{\omega _{2}}(1)\Phi _{j3,m_{3},\overline{m}%
_{3}}^{\omega _{3}}(\infty )\right\rangle _{WZW}\sim \prod_{i=1}^{3}\frac{%
\Gamma (-m_{i}-j_{i})}{\Gamma (j_{i}+1+\bar{m}_{i})}\prod_{t=1}^{s}\int
d^{2}v_{t}\left\langle e^{\sqrt{\frac{2}{k-2}}(j_{1}+1)\widehat{\varphi }%
(0)}\right. \times
\end{equation*}%
\begin{equation*}
\times e^{\sqrt{\frac{2}{k-2}}(j_{2}+1)\widehat{\varphi }(1)}e^{\sqrt{\frac{2%
}{k-2}}(j_{3}+1)\widehat{\varphi }(\infty )}\prod_{t=1}^{s}\left. e^{\sqrt{%
\frac{2}{k-2}}\widehat{\varphi }(v_{t})}\right\rangle _{S_{L}[\mu =0]}\delta
\left( s+j_{1}+j_{2}+j_{3}+1+\frac{k}{2}\right) ,
\end{equation*}%
and, up to an irrelevant $k$-dependent ($j$-$m$-independent) factor and
having fixed the value of the black hole mass, the final result reads%
\begin{equation*}
\left\langle \Phi _{j_{1},m_{1},\overline{m}_{1}}^{\omega _{1}}(0)\Phi
_{j_{2},m_{2},\overline{m}_{2}}^{\omega _{2}}(1)\Phi _{j3,m_{3},\overline{m}%
_{3}}^{\omega _{3}}(\infty )\right\rangle _{WZW}=\left( \pi \gamma \left( 
\frac{1}{k-2}\right) \right) ^{-j_{1}-j_{2}-j_{3}-\frac{k}{2}%
-1}\prod_{i=1}^{3}\frac{\Gamma (-m_{i}-j_{i})}{\Gamma (j_{i}+1+\bar{m}_{i})}%
\times
\end{equation*}%
\begin{eqnarray}
&&\times \frac{G_{k}(j_{1}+j_{2}+j_{3}+\frac{k}{2})G_{k}(-j_{1}-j_{2}+j_{3}-%
\frac{k}{2})G_{k}(j_{1}-j_{2}-j_{3}-\frac{k}{2})G_{k}(1+j_{1}-j_{2}+j_{3}-%
\frac{k}{2})}{\gamma \left( -j_{1}-j_{2}-j_{3}-\frac{k}{2}\right) \gamma
\left( -\frac{2j_{2}+1}{k-2}\right)
G_{k}(-1)G_{k}(2j_{1}+1)G_{k}(1-k-2j_{2})G_{k}(2j_{3}+1)}\times  \notag \\
&&\times \delta (m_{1}+m_{2}+m_{3}-k/2)\delta (\bar{m}_{1}+\bar{m}_{2}+\bar{m%
}_{3}-k/2)\delta (s+j_{1}+j_{2}+j_{3}+1+k/2).  \label{result11}
\end{eqnarray}%
where the special function $G_{k}(x)$ is defined through%
\begin{equation*}
G_{k}(x)=(k-2)^{\frac{x(k-1-x)}{2(k-2)}}\Gamma _{2}(-x|1,k-2)\Gamma
_{2}(k-1+x|1,k-2),
\end{equation*}%
in terms of the Barnes function $\Gamma _{2}(x|1,y)$%
\begin{equation*}
\log \Gamma _{2}(x|1,y)=\lim_{\varepsilon \rightarrow 0}\frac{d}{%
d\varepsilon }\sum\limits_{n=0}^{\infty }\sum\limits_{m=0}^{\infty }\left(
(x+n+my)^{-\varepsilon }-(1-\delta _{n,0}\delta _{m,0})(n+my)^{-\varepsilon
}\right) ,
\end{equation*}%
where the presence of the factor $(1-\delta _{n,0}\delta _{m,0})$ in the
r.h.s. means that the sum in the second term does not take into account the
step $m=n=0$. Expression (\ref{result11}) does reproduce the exact result,
so that agrees with the result obtained in Refs. \cite{FZZ,MO3,YoPLB2005}.
The details of the computation can be found in \cite{gaston}. Rather than an
application, the calculation of (\ref{result11}) can be considered as a
consistency check of the representation (\ref{cardinate}) proposed here (and
in \cite{0511252})\ to represent WZW correlators. Besides, it also
represents an operative advantage since (unlike other free field
realizations for which the computation of violating winding three-point
function involves the additional spectral flow operator) this turns out to
be integrable in terms of the Dotsenko-Fateev type integrals (cf. the
calculations in Refs. \cite{GN3,MO3}). Nevertheless, it is worth pointing
out that the consistency check discussed here is the most simple
(non-trivial) computation one can do within this framework; this is because
it did not involve the degenerate Liouville fields $V_{-1/2b}$. Unlike, the
screening that we did use to realize (\ref{result11}) was the new one we
introduced; namely, the operator $\mathcal{T}_{1-k,k,k}=e^{\sqrt{\frac{2}{k-2%
}}\widehat{\varphi }}=e^{-\sqrt{\frac{2}{k-2}}(k-1)\varphi +i\sqrt{2k}X}$,
which represents a $n=2$ perturbation. A less trivial consistency check
would be that of trying to reproduce the winding-conservative WZW
three-point function in the often called $m$-basis, which would require to
make use of a non-trivial integral representation of the (hypergeometric)
special function of the kind studied in Refs. \cite{Satoh,HS,Lore}. Related
to this point, let us mention that explicit expressions for Liouville
four-point functions involving one degenerate state $V_{-\frac{1}{2b}}$ were
recently obtained \cite{Fateev2,BelavinZamolodchikov}. According to the
relation (\ref{rrtt}), these four-point functions are the ones representing
three-point functions that conservate the winding conservation in the WZW
side.

Other applications of the Stoyanovsky-Ribault-Teschner correspondence (\ref%
{rrtt}), (\ref{duality}), were early discussed in Ref. \cite%
{Apl1,Apl2,YoYu,Takayanagi}. In subsection 4.5.3, we will review one of the
observations made in \cite{YoYu}.


\subsection{Remarks}

\subsubsection{A comment on generalized minimal gravity}

Now, we would like to make a brief comment on the theory defined by the
action (\ref{SA}) and the perturbation (\ref{SB}); and let us focus our
attention on the two-dimensional sector corresponding to the fields $%
\widehat{\varphi }$ and $\widehat{X}$. Because of the field redefinitions (%
\ref{newc}), it turns out that the theory could be written as the Liouville
theory coupled to a $c<1$ CFT. Then, the natural question arises as to
whether such a $c<1$ model can be identified with one of the quoted minimal
models. As it is well known, the CFT minimal models are characterized by two
integers $p$ and $q$ which yield the value of the central charge, being $%
c=1-6(\beta ^{-1}-\beta )^{2}$ with rational\footnote{%
Besides, a generalized version of these CFTs can be considered, being valid
for generic values of $\beta ,$ \cite{Zreloaded}.} $\beta ^{2}=p/q$
satisfying $q>p$ (so that $\beta <1$). In our case, the value of the central
charge of the $c<1$ theory (corresponding to the part of the theory governed
by the field $\widehat{X}$) turns out to be $c=1-6k$ and, then, in order to
identify this with one of the minimal models we should demand $%
k=(p-q)^{2}/pq $ (that is $k=\left( \beta ^{-1}-\beta \right) ^{2}$) \cite%
{BPZ}. However, since we are interested in the whole range $k>2$, it turns
out that the condition $c=1-6(p-q)^{2}/pq$ is only consistent with
particular values of $k$. One example is precisely the model ($p=1,$ $q=4$)
which does correspond to $k=9/4,$ which is the value of $k$ for the 2D
theory on the coset. In such case, and taking into account that $k$ also
satisfies $k=2+b^{-2},$ we would have $\beta =b$ so that the theory
corresponds to the often called 2D minimal gravity (model that is supposed
to be exactly solved). For more general case, the 2D theory defined by the
fields $\widehat{\varphi }$ and $\widehat{X}$ can be regarded as the
Liouville theory coupled to a generalized minimal model (with
non-necessarily rational $\beta ^{2}$) perturbed by (\ref{SB}). Such
perturbation would then correspond to a Liouville-dressed operator in the
minimal model too. The operators of the minimal models admit a
representation in terms of the exponential form $\Phi _{mn}=e^{i\alpha _{mn}%
\widehat{X}},$ having conformal dimension $h_{mn}=\frac{1}{4}(m\beta
^{-1}-n\beta )^{2}-\frac{1}{4}(\beta ^{-1}-\beta )^{2}=\alpha _{mn}(\alpha
_{mn}+\beta -\beta ^{-1})$ for two positive integers $m$ and $n$; that is,
the momenta can take values $\alpha _{mn}=\frac{1}{2}(n-1)\beta -\frac{1}{2}%
(m-1)\beta ^{-1}$ or $\alpha _{mn}=\frac{1}{2}(m+1)\beta ^{-1}-\frac{1}{2}%
(n+1)\beta $. So, a perturbation operator with the form $e^{i\sqrt{k/2}n%
\widehat{X}+\sqrt{2}a_{n}\widehat{\varphi }}$ can be regarded as a dressed
field $\Phi _{n-1,n-1}$ of the minimal model ($p,q$) with $k=(p-q)^{2}/pq$.
In these terms, what we have proven is a correspondence between $N$-point
functions in the WZW theory and a subset of correlation functions of
perturbed Liouville gravity coupled to generalized minimal models.

\subsubsection{Duality between tachyon-like backgrounds}

By using the relation between Liouville correlators and WZW correlators we
wrote down identity (\ref{duality}). This gives a dual description for the
2D string theory in the black hole background. One of the questions that
arise is about the relation between (\ref{duality}) and the standard FZZ
correspondence. In fact, both models appear as alternative dual descriptions
of the WZW theory, so that we can use WZW correlators as an intermediate
step to eventually write the following seemly self-duality relation%
\begin{equation*}
\prod_{r=1}^{s_{+}}\int d^{2}u_{r}\prod_{t=1}^{s_{-}}\int
d^{2}v_{t}\left\langle \prod_{i=1}^{N}\mathcal{T}_{j_{i},m_{i},\overline{m}%
_{i}}(z_{i})\prod_{r=1}^{s_{+}}\mathcal{T}_{1-\frac{k}{2},-\frac{k}{2},\frac{%
k}{2}}(u_{r})\prod_{t=1}^{s_{-}}\mathcal{T}_{1-\frac{k}{2},\frac{k}{2},-%
\frac{k}{2}}(v_{t})\right\rangle _{S_{[\lambda =0]}}\sim
\end{equation*}%
\begin{equation}
\sim \prod_{r=1}^{\widetilde{s}_{+}}\int d^{2}u_{r}\prod_{t=1}^{\widetilde{s}%
_{++}}\int d^{2}\omega _{t}\left\langle \prod_{i=1}^{N}\widetilde{\mathcal{T}%
}_{j_{i},m_{i},\overline{m}_{i}}(z_{i})\prod_{r=1}^{\widetilde{s}_{+}}%
\mathcal{T}_{1-\frac{k}{2},\frac{k}{2},\frac{k}{2}}(u_{r})\prod_{t=1}^{%
\widetilde{s}_{++}}\mathcal{T}_{1-k,k,k}(\omega _{t})\right\rangle
_{S_{[\lambda =0]}},  \label{selfduality}
\end{equation}%
where the fearful symbol $\sim $ stands to make explicit the fact that this
identity depends on the details of how the FZZ conjecture relates the WZW
correlators to those of sine-Liouville theory\footnote{%
As far as we know, the checks of FZZ duality were performed by comparing the
analytic structures of both theories rather than verifying exact numerical
matching.}. This relation between correlators, realized by means of the
Coulomb gas realization, yields a non-trivial integral identity. On the
other hand, one can wonder whether (\ref{selfduality}) has to be referred as
a self-duality of sine-Liouville field theory or not. In fact, it merely
looks like a duality between two different deformations of the linear
dilaton theory (\ref{S0}) rather than a \textquotedblleft self-duality".
However, one can see that both sides in the identity above are in some sense
connected to sine-Liouville theory, and not only the left hand side.
Actually, the perturbation $\mathcal{T}_{1-k,k,k}$, that represents the
momentum $n=2$ operator, is connected to that of $n=1$ by the conjugation
relation (\ref{conjugate}). That is, while $j=1-m=1-k$ for the $n=2$
operator $\mathcal{T}_{1-k,k,k}$, the dual momenta (dual according to (\ref%
{j})) are\footnote{%
Strictly speaking, one has to consider the automorphism $m\rightarrow 
\widetilde{m}=-jk-m(k-1)-k/2$ instead of (\ref{j}), which is a composition
with the reflection $m\rightarrow -m$.} $\widetilde{j}=1+\widetilde{m}=1-k/2$%
, and correspond to the momenta of the $n=-1$ operator $\mathcal{T}_{1-\frac{%
k}{2},-\frac{k}{2},-\frac{k}{2}}$. Thus, we could relate the correlators in
the l.h.s. of (\ref{selfduality}) to the following one\footnote{%
up to a $k$-dependent factor of the form $\left( b_{k}\right) ^{\widetilde{s}%
_{-}}$, with $b_{k}$ being independent on $j_{i},m_{i}$ and $\overline{m}%
_{i} $.}%
\begin{equation}
\sim \prod_{r=1}^{\widetilde{s}_{+}}\int d^{2}u_{r}\prod_{t=1}^{\widetilde{s}%
_{-}}\int d^{2}\omega _{t}\left\langle \prod_{i=1}^{N}\widetilde{\mathcal{T}}%
_{j_{i},m_{i},\overline{m}_{i}}(z_{i})\prod_{r=1}^{\widetilde{s}_{+}}%
\mathcal{T}_{1-\frac{k}{2},+\frac{k}{2},+\frac{k}{2}}(u_{r})\prod_{t=1}^{%
\widetilde{s}_{-}}\widetilde{\mathcal{T}}_{1-\frac{k}{2},-\frac{k}{2},-\frac{%
k}{2}}(\omega _{t})\right\rangle _{S_{[\lambda =0]}}  \label{uhh}
\end{equation}%
with $\frac{k-2}{2}\widetilde{s}_{+}-\widetilde{s}_{-}-\frac{k}{2}\left(
N-2\right) =\frac{k-2}{2}\left( s_{+}+s_{-}\right) $ and $\widetilde{s}%
_{+}-(N-2)=s_{+}-s_{-}$. Hence, (\ref{selfduality}) turns out to be a
twisted version of the sine-Liouville model, i.e. can be written as in (\ref%
{uhh}). The presence of the tildes $\sim $ on the operators in (\ref{uhh})
gives rise to the expression \textquotedblleft twisted"; twisted in the
sense that (\ref{j}) is applied to the operators $\mathcal{T}_{1-\frac{k}{2}%
,-\frac{k}{2},-\frac{k}{2}}$ but is not applied to the operators $\mathcal{T}%
_{1-\frac{k}{2},+\frac{k}{2},+\frac{k}{2}}$. This kind of relation between
correlators (\ref{selfduality}) and (\ref{uhh}) is reminiscent of what
happens in the WZW theory, where standard and conjugate representations
stand as alternative realizations of the same correlation functions. Thus,
this suggests that (\ref{selfduality}) could be manifesting some kind of
self-duality relating two different realization of the same conformal theory%
\footnote{%
Let us also mention that another realization of the same correlators is
possible if one replace $\widetilde{\mathcal{T}}_{1-\frac{k}{2},-\frac{k}{2}%
,-\frac{k}{2}}\propto e^{-\sqrt{\frac{2}{k-2}}(k-1)\widehat{\varphi }+i\sqrt{%
2k}\widehat{X}}$ by its $k-2$ power $e^{-\sqrt{2(k-2)}(k-1)\widehat{\varphi }%
+i\sqrt{2k}(k-2)\widehat{X}}$. This is because of the Liouville self-duality
under $b\leftrightarrow b^{-1}$.}. Morally, the price to be paid to twist
(namely, to conjugate) the $N$ vertex operators $\mathcal{T}_{j_{i},m_{i},%
\overline{m}_{i}}$ in (\ref{selfduality}) is that of twisting the
left-handed screening operators $\mathcal{T}_{1-\frac{k}{2},-\frac{k}{2},-%
\frac{k}{2}}\rightarrow \widetilde{\mathcal{T}}_{1-\frac{k}{2},-\frac{k}{2},-%
\frac{k}{2}}\propto \mathcal{T}_{1-k,k,k}$, while keeping the right-handed $%
\mathcal{T}_{1-\frac{k}{2},+\frac{k}{2},+\frac{k}{2}}$ unchanged.
Consequently, the number of insertions changes from $s_{-}$ to $\widetilde{s}%
_{++}=\widetilde{s}_{-}$ (and also from $s_{+}$ to $\widetilde{s}_{+}$) by
keeping the formal relation $N-2=\widetilde{s}_{+}-s_{+}+s_{-}$ fixed.
Roughly speaking, the right hand side of (\ref{selfduality}) looks like a
\textquotedblleft half" of a sine-Liouville theory, because just one of the
two exponential operators $\mathcal{T}_{1-\frac{k}{2},\pm \frac{k}{2},\pm 
\frac{k}{2}}$ that form the cosine interaction (\ref{cos2}) is present,
while the operators $\mathcal{T}_{1-k,k,k}$ seem to arise there for
compensating the conservation laws that make the correlator to be nonzero.
In the WZW theory, the analogue to the \textquotedblleft twisting" that
connects the operators $\widetilde{\mathcal{T}}_{j,m,\overline{m}}$ to
operators $\mathcal{T}_{j,m,\overline{m}}$ would be the relation existing
between conjugate and standard representations of the $\widehat{sl}(2)_{k}$
vertex algebra \cite{DVV,BK,GN2,GN3,GL}. The relation between
representations $\widetilde{\mathcal{T}}_{j,m,\overline{m}}$ and $\mathcal{T}%
_{j,m,\overline{m}}$ connects operators of the winding sector $n$ to those
of the sector $n+1$. Presumably, the twisted version of the FZZ duality we
presented in (\ref{duality}) can be extended in order to include higher
momentum and winding modes $n>2$. This would rise the obvious question as to
what would these twisted sectors be describing in terms of the black hole
picture. As it was pointed out in \cite{KKK}, if the $c=1$ theory is
perturbed by operators of the sector $n,$ then it behaves equivalently to
the theory compactified in a different\ radius $R/n$ and perturbed by the
sine-Liouville operators. In some sense, this is related to what was early
studied in Ref. \cite{Mukhi}. Nevertheless, the perturbation we considered
here presents operators of both sectors $n=1$ and $n=2$, so being a sort of
chirally twisted case. We would like to understand this deformations better.
Our hope is to make contact to the results of Refs. \cite{MMP} and \cite%
{Mukhi} in trying to answer this question, but this certainly requires
further study.

\subsubsection{The \thinspace $c\rightarrow 0$ limit of the $Liouville\times
U(1)\times \mathbb{R}$ model}

To conclude, we would like to discuss the particular limit where the central
charge of the model (\ref{cardinate}) vanishes. This was first studied in
Refs. \cite{Takayanagi2}\ and \cite{YoYu} (see also \cite{Nichols}-\cite%
{Nichols2}). This limit corresponds to $k\rightarrow 0,$ which, in fact, is
far from being well understood. Actually, one can rise several question
concerning whether in such a limit the CFT is well defined or not. However,
let us avoid these questions here and merely assume that such an extension
is admissible. In the limit $k\rightarrow 0$, the Liouville central charge
becomes $c_{L}=-2$ while the background charge for the field $\widehat{X}$
vanishes, so that the central charge for the $U(1)\times \mathbb{R}$ theory
(i.e. the fields $\widehat{X}$ and $T $) turns out to be $+2$. The
functional form of the correlation functions in the $k\rightarrow 0$ limit
requires a careful analysis because of subtle features arising through the
analytic continuation in the $b$ complex plane \cite{SchomerusII,Zreloaded}.
However, we can further speculate and assume for a while that an extension
of the correspondence (\ref{rrtt}) between WZW and Liouville theory still
holds at $k=0$. At this point, the sine-Liouville action actually coincides
with the Liouville action supplemented with that of a $c=1$ field $\widehat{X%
}$. This is because of the identification $b^{-2}=k-2$ and the fact that $Q=
- \widehat{Q}$ at the point $k=0$. Besides, at $k=0$ the sine-Liouville
interaction (\ref{cos}) does correspond to the Liouville cosmological
constant $e^{i\widehat{\varphi }/\sqrt{2}}$. This suggestive matching
between both actions can be tested at the level of correlation functions as
well. In fact, with the authors of \cite{Takayanagi2}, we could assume that
the FZZ conjecture is still valid in the limit $k\rightarrow 0$ and, then,
by invoking the Ribault-Teschner formula (\ref{rrtt}), eventually conclude
that the sine-Liouville correlators model coincide with the correlators of
the Liouville theory (times the free boson $\widehat{X}$) at $k=0$. To see
this, let us point out the following remarkable facts: First, notice that,
because we are taking a limit $R=\sqrt{k/2}$ going to zero (\textit{i.e.}
the asymptotic radius of the cigar), it is just enough to observe what
happens with the modes $m=\bar{m}=0$ on the cigar. From the point of view of
the T-dual model, the dual radius $\tilde{R}\sim 1/\sqrt{k}$ of the cylinder
goes to infinity and the states with finite momentum $p=\frac{m}{\sqrt{k}}$
(keeping $p$ fixed) decouple generating a $U(1)$ factor $\sim e^{i\sqrt{2}p%
\widehat{X}}$ in the correlation functions. Secondly, one can show (see \cite%
{YoYu}) that for $k=0$ the formula (\ref{rrtt}) reads 
\begin{equation}
\left\langle \Phi _{j_{1},m_{1},\overline{m}_{1}}^{\omega
_{1}}(z_{1})...\Phi _{j_{N},m_{N},\overline{m}_{N}}^{\omega
_{N}}(z_{N})\right\rangle _{WZW}\sim \prod_{i=1}^{N}{\mathcal{R}}%
_{0}(j_{i},0)\ \langle V_{-\frac{i}{\sqrt{2}}j_{1}}(z_{1})...V_{-\frac{i}{%
\sqrt{2}}j_{N}}(z_{N})\rangle _{S_{L}[\mu ]};  \label{pros}
\end{equation}%
with $p_{1}+p_{2}+...p_{N}=\omega _{1}+\omega _{2}+...\omega _{N}=M-N+2=0$.
The function ${\mathcal{R}}_{k}(j,m)$ is the reflection coefficient of WZW
model, which is given by the two-point function (\ref{FiRulete2}). The
arising of these reflection coefficients (one for each vertex operator) is
ultimately attributed to the fact that the momenta of the WZW vertex
operators were the Weyl reflected $\widehat{j}_{i}=-1-j_{i}$ instead of $%
j_{i}$ (notice that the Liouville correlator in (\ref{pros}) scales like $%
\mu ^{\hat{j}_1+...\hat{j}_N+1}$). We also observe in (\ref{pros}) that,
besides the $s$ integrals over the screening insertions required in the
Liouville correlators, we implicitly have $M=N-2$ additional integrals over
the variables $v_{t}$ where $M$ operators $V_{-1/2b}(v_{t})$ are inserted.
This is consistent with what one would expect since $k=0$ implies $%
b^{2}=-1/2 $ and then the degenerate fields $V_{-1/2b}$ turn out to agree
with the screening operators $V_{b}$. Hence, at $k=0$ the integrals over
such variables $v_{t}$ are nothing more than screening insertions in
Liouville correlation functions\footnote{%
G.G. specially thanks Yu Nakayama for collaboration in this particular
computation. See Ref. \cite{YoYu}.}, and this is the reason why we did not
explicitly write them in (\ref{pros}). This shows that the Ribault-Teschner
formula turns out to be consistent with the FZZ conjecture. That is, at $k=0$
sine-Liouville agrees with the product between Liouville theory and a free $%
c=1$ boson, so that for the particular case $k=0$ equation (\ref{rrtt})
actually states the identity between $N$-point correlation functions in
sine-Liouville theory and $N$-point correlation functions in the 2D black
hole. Nevertheless, we should emphasize that all these digressions are
strongly based on the assumption that the CFT is still well defined in the
regime $k<2$ and, as far as we know, this is still far from being clear.

\section{Conclusions}

It is usually accepted that, probably, the FZZ duality is just an example of
a more general phenomenon which should be interesting to understand in a
deeper way \cite{KKK,MMP}. The purpose of this paper was precisely to
discuss an example of such kind of generalization. We studied a
correspondence between two-dimensional string theory in the euclidean black
hole ($\times time$) and a (higher mode) tachyon perturbation of a linear
dilaton background. Our main result is presented in Eq. (\ref{duality}).

The tachyon perturbation we considered here corresponds to momentum modes $%
n=1$ and $n=2,$ and so it can be considered as a kind of deformation of the
standard FZZ sine-Liouville theory. We argued that such a \textquotedblleft
deformation" (or \textquotedblleft twisting" in the sense of (\ref{j})) can
be thought of as a conjugate representation of the sine-Liouville
interaction term, presumably related to the conjugate representations of
operators in the WZW model \cite{GL}.

In section 4 we have given a dictionary that permits to express any $N$%
-point correlation function in the $SL(2,\mathbb{R})_{k}$ WZW model on the
sphere topology in terms of a correlation function in the tachyon perturbed
linear dilaton background, and we have given a precise prescription for
computing those correlators in the Coulomb gas approach. This correspondence
between correlators was proven by rewriting a nice formula worked out by S.
Ribault and J. Teschner in Refs. \cite{R,RT}, which directly follows from
the relation between the solutions of the KZ and the BPZ equations. Our
result (\ref{yesto}) realizes the general version of the formula proven in 
\cite{R}. In fact, following \cite{0511252}, we showed that the auxiliary
overall function $F_{k}(z_{1},...z_{N};w_{1},...w_{M})$ standing in the
Ribault-Teschner formula (\ref{rrtt}) can be also seen as coming from the
correlation functions of a linear dilaton CFT perturbed by a tachyon-like
operator of higher ($n\geq 1$) momentum modes. Thus, the twisted dual we
discussed here turns out to be a free field realization of the
Ribault-Teschner formula. A remarkable feature of such realization is that
the $n=2$ mode perturbation $\mathcal{T}_{1-k,k,k}$ turns out to be related
to the sine-Liouville potential in the same way as to how the twisted
tachyon-like vertex operators $\widetilde{\mathcal{T}}_{j_{i},m_{i},%
\overline{m}_{i}}$ are related to the operators $\mathcal{T}_{j_{i},m_{i},%
\overline{m}_{i}}$ of the standard FZZ prescription. Both representations $%
\widetilde{\mathcal{T}}_{j_{i},m_{i},\overline{m}_{i}}$ and $\mathcal{T}%
_{j_{i},m_{i},\overline{m}_{i}}$ have the same eigenvalues under the Cartan $%
U(1)$ generator $J_{0}^{3}$ and the Virasoro-Casimir operator $L_{0}$.
Besides, it turns out that the fact that $\mathcal{T}_{j_{i},m_{i},\overline{%
m}_{i}}$ and $\widetilde{\mathcal{T}}_{j_{i},m_{i},\overline{m}_{i}}$
transform distinctly under the action of the $\widehat{sl}(2)_{k}$
generators $J_{n}^{\pm }$ combines with the fact that the $n=2$ operator $%
\mathcal{T}_{1-k,k,k}$ transforms non-trivially under those generators
either, and this makes the correlation functions (\ref{yesto})\ behave
properly under the $sl(2)_{k}$ algebra.

Tachyon-like perturbations of the linear dilaton background involving higher
winding modes were also studied recently by Mukherjee, Mukhi and Pakman in
Ref. \cite{MMP}, where they presented a generalized perspective of the FZZ
correspondence. One of the task for the future is to understand the relation
to \cite{MMP} better. Besides, the understanding of the connection of our
result to the standard FZZ correspondence also deserves more analysis.
Regarding this, we would like to conclude by mentioning that the idea of the
proof of (\ref{duality}) given in section 4 here could be actually adapted
to prove the standard FZZ duality (on the sphere) if one considers the
appropriated pieces in the literature. A key point in doing this would be a
result obtained some time ago by V. Fateev, who has found a very direct way
of showing the relation between correlation functions in both Liouville and
sine-Liouville theories \cite{Fateev}. Such connection, once combined with
the Ribault-Teschner formula \cite{RT,R}, would yield a proof of the FZZ
duality at the level of correlation functions on the sphere topology without
resorting to arguments based on supersymmetry.

\bigskip

This paper is an extended version of the authors' contribution to the XVIth
International Colloquium on Integrable System and Quantum Symmetries, held
in Prague, in June 2007. A brief version was published in Rep. Math. Phys. 
\textbf{61}.2 (2008) 151-162. Besides, these notes are based on Refs. \cite%
{0511252,gaston}\ and summarize the contents of the seminars that one of the
authors has delivered at several institutions in the last year and a half.
G.G. would like to thank S. Murthy, Yu Nakayama, K. Narain, A. Pakman, S.
Ribault and V. Schomerus for conversations, for e-mail exchanges, and for
very important comments. He is also grateful to V. Fateev for sharing his
unpublished work \cite{Fateev}. The partial support of Universidad de Buenos
Aires, Agencia ANPCyT, and CONICET through grants UBACyT X861, PICT 34557,
PIP6160 is also acknowledged.

\bigskip

\textbf{Note:} After our paper appeared in arXives, the higher genus
generalization of the Ribault-Teschner formula was done in the very nice
paper \cite{HikidaSchomerus1}. There, Hikida and Schomerus derived a
generalization of (\ref{rrtt}) by using the path integral approach. One of
the fabulous applications of Hikida-Schomerus formula was that of giving a
proof of the FZZ duality conjecture \cite{HikidaSchomerus2}. A crucial step
in the proof giving in \cite{HikidaSchomerus2} is the result of Ref. \cite%
{0511252}, which we discussed here within a similar context.

The authors of \cite{HikidaSchomerus2} also asserted that the twisted dual
model described by the action (\ref{S}) perturbed with operator (\ref%
{Pupapupapupa56}) is actually not well defined. We understand they meant
that such an action cannot be taken literally for all purposes as it is not
weakly coupled. This is of course true, and this is why we emphasized in
subsection 4.2 that the twisted model we presented here makes sense only
when the appropriate prescription for computing correlattion functions is
considered. It is well known that free field realization and Coulomb
gas-like prescriptions work well in conformal models even when the
interaction terms are not necessarily well-defined perturbatively. This is,
for instance, the case of Liouville theory with the \textit{dual} screening
charge, and this is also the case of sine-Liouville theory, where the
Coulomb gas prescription is known to be suitable for computing three-point
correlators. Then, we understand the authors of \cite{HikidaSchomerus2}
would agree in that our twisted model can be consistently used to compute
correlation functions, even though writing its action could be misleading if
taken literally. After all, if appropriately used, this represents a free
field realization of identity (\ref{rrtt}).

G.G. thanks V. Schomerus for a conversation about this point.

\end{document}